\documentclass[journal]{IEEEtran}

\usepackage{ifpdf}
\usepackage{url}
\ifpdf
\else
\fi

\usepackage{cite}
\usepackage{balance}
\usepackage{bm,comment,color}

\ifCLASSINFOpdf
  \usepackage[pdftex]{graphicx}
  \graphicspath{{../pdf/}{../jpeg/}}
  \DeclareGraphicsExtensions{.pdf,.jpeg,.png}
\else
  \usepackage[dvips]{graphicx}
  \graphicspath{{../eps/}}
  \DeclareGraphicsExtensions{.eps}
\fi
\usepackage{amsmath}

\usepackage{algorithm}
\usepackage{algpseudocode}
\algtext*{EndWhile}
\algtext*{EndIf}
\algtext*{EndFor}

\usepackage{array}
\usepackage{amsfonts} 
\usepackage{amssymb}
\usepackage{esint} 
\usepackage{units}
\usepackage{multirow}
\usepackage{comment}

\usepackage{color}
\newcommand{\red}[1]{{\color{red}{#1}}} 
\newcommand{\blue}[1]{{\color{blue}{#1}}} 
 
\newcommand{\green}[1]{{\color{green}{#1}}}

\usepackage{booktabs}
\usepackage{pgfplots}
\usepackage{tikz}
\usetikzlibrary{calc}
\makeatletter
\newcommand{\gettikzxy}[3]{%
  \tikz@scan@one@point\pgfutil@firstofone#1\relax
  \edef#2{\the\pgf@x}%
  \edef#3{\the\pgf@y}%
}
\usetikzlibrary{spy,backgrounds}
\pgfplotsset{compat=newest}
\usetikzlibrary{plotmarks}
\usetikzlibrary{arrows.meta}
\usepgfplotslibrary{patchplots}
\usepackage{grffile}
\newlength\fheight 
\newlength\fwidth 
\usepgfplotslibrary{fillbetween}

\usepackage{acronym}

\usepackage{tabu,longtable}
\usepackage{subfig}
\usepackage{stfloats}
\usepackage[hidelinks]{hyperref}
\usepackage{xcolor}
\acrodef{3gpp}[3GPP]{3rd Gneration Partnership Project}
\acrodef{6d}[6D]{six-dimensional}
\acrodef{ad}[AD]{autonomous drive}
\acrodef{adas}[ADAS]{advanced driver assistance system}
\acrodef{aoa}[AOA]{angles-of-arrival}
\acrodef{aod}[AOD]{angles-of-departure}
\acrodef{aosa}[AOSA]{array-of-subarray}
\acrodef{ap}[AP]{access point}
\acrodef{bs}[BS]{base station}
\acrodef{bse}[BSE]{beam squint effect}
\acrodef{cdf}[CDF]{cumulative distribution function}
\acrodef{cir}[CIR]{channel impulse response}
\acrodef{csi}[CSI]{channel state information}
\acrodef{coa}[COA]{curvature of arrival}
\acrodef{crb}[CRB]{Cram\'er-Rao bound}
\acrodef{ccrb}[CCRB]{constrained Cram\'er-Rao bound}
\acrodef{icrb}[ICRB]{intrinsic Cram\'er-Rao bound}
\acrodef{dof}[DOF]{degrees of freedom}

\acrodef{dbscan}[DBSCAN]{density-based spatial clustering of applications with noise}
\acrodef{elaa}[ELAA]{extremely-large antenna array}
\acrodef{ekf}[EKF]{extended Kalman filter}
\acrodef{phd}[PHD]{probability hypothesis density}
\acrodef{ff}[FF]{far field}
\acrodef{fim}[FIM]{Fisher information matrix}
\acrodef{gnss}[GNSS]{global navigation satellite system}
\acrodef{gps}[GPS]{global positioning system}
\acrodef{icrb}[ICRB]{Intrinsic Cram\'er-Rao bound}
\acrodef{imu}[IMU]{inertial measurement unit}
\acrodef{ip}[IP]{incidence point}
\acrodef{kld}[KLD]{Kullback–Leibler divergence}
\acrodef{las}[L\&S]{localization and sensing}
\acrodef{los}[LOS]{line-of-sight}
\acrodef{mae}[MAE]{mean absolute value}
\acrodef{map}[MAP]{multipath-assisted positioning}
\acrodef{mimo}[MIMO]{multiple-input-multiple-output}
\acrodef{siso}[SISO]{single-input-single-output}
\acrodef{mle}[MLE]{maximum likelihood estimator}
\acrodef{mm}[MM]{mismatched model}
\acrodef{mpc}[MPC]{multipath component}
\acrodef{nlos}[NLOS]{non-line-of-sight}
\acrodef{nf}[NF]{near field}
\acrodef{nr}[NR]{new radio}
\acrodef{ofdm}[OFDM]{orthogonal frequency division multiplexing}
\acrodef{pbd}[PBD]{partial blockage detection}
\acrodef{prs}[PRS]{positioning reference signal}
\acrodef{psd}[PSD]{power spectral density}
\acrodef{pss}[PSS]{primary synchronization signal}
\acrodef{mse}[MSE]{mean squared error}
\acrodef{rmse}[RMSE]{root mean squared error}
\acrodef{rf}[RF]{radio frequency}
\acrodef{rfc}[RFC]{radio frequency chain}

\acrodef{ris}[RIS]{reconfigurable intelligent surface}
\acrodef{rss}[RSS]{received signal strength}
\acrodef{rtk}[RTK]{real-time kinematic}
\acrodef{rtt}[RTT]{round-trip-time}
\acrodef{sa}[SA]{sub-array}
\acrodef{simo}[SIMO]{single-input-multiple-output}
\acrodef{slam}[SLAM]{simultaneous localization and mapping}
\acrodef{snr}[SNR]{signal-to-noise ratio}
\acrodef{sns}[SNS]{spatial non-stationarity}
\acrodef{sp}[SP]{scattering point}
\acrodef{ssb}[SSB]{synchronization signal/physical broadcast channel block}
\acrodef{swm}[SWM]{spherical wave model}
\acrodef{tdd}[TDD]{time division duplex}
\acrodef{tdoa}[TDOA]{time-difference-of-arrival}
\acrodef{tm}[TM]{true model}
\acrodef{toa}[TOA]{time-of-arrival}
\acrodef{ue}[UE]{user equipment}
\acrodef{upa}[UPA]{uniform planar array}
\acrodef{ura}[URA]{uniform rectangular array}
\acrodef{va}[VA]{virtual anchor}
\acrodef{5g}[5G]{fifth-generation communication system}

\acrodef{dsrd}[DSRD]{diffuse-specular reflection disentanglable}
\acrodef{dr}[DR]{diffuse reflection}
\acrodef{sr}[SR]{specular reflection}
\acrodef{wss}[WSS]{wide-sense stationary}
\acrodef{fp}[FP]{first path}
\acrodef{lvm}[LVM]{latent variable model}
\acrodef{elbo}[ELBO]{evidence lower bound}
\acrodef{isac}[ISAC]{Integrated Sensing and Communication}

\long\def\comment#1{}

\newfont{\bbb}{msbm10 scaled 700}

\newfont{\bb}{msbm10 scaled 1100}

\begin{document}

\title{MudiNet: Task-guided Disentangled Representation Learning for 5G Indoor Multipath-assisted Positioning}

\author{Ye~Tian,
Xueting~Xu
and~Ao~Peng*,~\IEEEmembership{Member,~IEEE},

\thanks{Y.~Tian is with the Institute of Industrial Science, The University of Tokyo, Tokyo 153-8505, Japan (E-mail: tianye\_ti@kmj.iis.u-tokyo.ac.jp)}

\thanks{X.~Xu is with the Department of Broadband Communications, Pengcheng Laboratory, Shenzhen 518055, China (E-mail: xuxueting4728@gmail.com)}

\thanks{A. Peng is with the School of Informatics, Xiamen University, Xiamen, Fujian 361005, China (Corresponding Author, E-mail: pa@xmu.edu.cn)}

}

\maketitle

\begin{abstract}
In the \ac{5g}, \ac{map} has emerged as a promising approach. With the enhancement of signal resolution, \ac{mpc} are no longer regarded as noise but rather as valuable information that can contribute to positioning. However, existing research often treats reflective surfaces as ideal reflectors, while being powerless in the face of indistinguishable multipath caused by diffuse reflectors. This study approaches diffuse reflectors from the perspective of uncertainty, investigating the statistical distribution characteristics of indoor diffuse and specular reflectors. Based on these insights, a task-guided disentangled representation learning method leveraging multi-time \ac{cir} observations is designed to directly map \ac{cir}s to positions, while mitigating the adverse effects of components that contribute minimally to localization accuracy (e.g., diffuse multipath).In this semi-supervised learning framework, a global feature extraction architecture based on self-attention is proposed to capture location-independent wireless environmental information, while an MLP is employed to extract the time-varying features related to \ac{ue} positions. Variational inference based on a \ac{lvm} is applied to separate independent features within the \ac{cir}, with position labels guiding the \ac{lvm} to express components more beneficial for localization. Additionally, we provide a feasibility proof for the separability of diffuse and specular environmental features in \ac{cir}s. Simulation results demonstrate that the proposed method achieves higher localization accuracy compared to conventional search-based localization methods, with enhanced robustness against indistinguishable multipath from diffuse reflectors.
\end{abstract}


\IEEEpeerreviewmaketitle
\acresetall 

\section{Introduction}
\label{sec:intro}
Indoor positioning is an essential technology for accurately locating objects or individuals within buildings where GNSS signals are weak or unavailable. This intricate challenge is met with a variety of technologies, including Wi-Fi\cite{shao2020accurate}, Bluetooth\cite{yu2021novel}, IMU\cite{wang2023recent}, UWB\cite{zhao2021new}, and notably, cellular networks, which are defined as a paradigm of \ac{isac}\cite{liu2022integrated}. These systems estimate positions based on such as signal strength, \ac{toa}, or \ac{aoa}. With the advent of \ac{5g}, cellular network positioning, leveraging wide bandwidth, dense networks, and advanced signal processing, has gained prominence due to its higher accuracy, lower latency, and broader coverage. This advancement is vital for applications in navigation, asset tracking, emergency response, and enhancing user experiences in smart buildings\cite{del2017survey}. 

In cellular network-based indoor positioning, \ac{mpc} of signals - reflections of walls, floors, and other obstacles - can significantly impact the accuracy of trilateration or triangulation, which are the most popular positioning methods. Traditionally viewed as a source of error, these \ac{mpc}s complicate the positioning process by affecting the determination of the \ac{los} path. However, as the resolution of the signal increases, recent advancements have turned this challenge into an opportunity through \ac{map}\cite{multipathsurvey2024}. Positioning accuracy can be enhanced by exploiting the channel geometric parameters of resolvable reflected signals. This innovative approach transforms what was once considered noise into valuable data, providing more reliable indoor positioning, especially in complex environments where \ac{los} signals are insufficient\cite{gentner2016multipath}. In certain scenarios, methods based on \ac{mpc} can achieve localization with even a single transmitter \cite{ge2021single,nazari2023mmwave}.

\ac{map} can be broadly categorized into non-geometric and geometric methods. Geometric methods, such as trilateral positioning and triangular positioning, utilize mathematical models to estimate positions based on the geometric parameter characteristics of \ac{mpc}s, the positioning accuracy is strictly dependent on the estimation precision of geometric parameters such as the \ac{toa} and \ac{aoa} of \ac{mpc}s. A single antenna can estimate the \ac{toa}\cite{liu2023multipath,liu2023machine} of \ac{mpc}s using peak tracking methods, while antenna arrays enable the estimation of \ac{aoa} of \ac{mpc}s with a certain level of precision and resolution \cite{zhou2022aoa,soltanaghaei2018multipath}. For geometric parameter estimation based on traditional signal processing methods, the phase information of specular reflection \ac{mpc}s is utilized, where an \ac{ekf} is employed to track the geometric parameters of \ac{mpc}s in \cite{li2019massive}. In \cite{du2024diffuse}, a \ac{phd} filter is employed to track complex \ac{mpc}s, where the distinction between specular reflection \ac{mpc}s and diffuse reflection \ac{mpc}s is discussed. The \ac{mpc}s caused by diffuse reflectors are treated as signals directly emitted from the diffuse reflectors. Data-driven approaches, such as using LSTM to mitigate the impact of NLOS environments on MPC estimation, can improve the accuracy of geometric parameter estimation for MPCs \cite{kim2022uwb}. Alternatively, variational inference has also been proposed in \cite{wang2024multipath, li2023variational} to achieve similar objectives. However, the accuracy of these geometric methods is often limited by their reliance on empirical models or approximations of \ac{mpc} propagation. After acquiring and tracking the \ac{mpc}s, the geometric method requires the second stage to complete the localization process. While these models are useful, each \ac{mpc} must be correctly associated with a \ac{va} to establish a complete geometric relationship for \ac{ue} position estimation. This introduces additional errors, namely association errors and VA estimation errors\cite{xu2023multi}. In addition, the information inherently contained in the waveform is overlooked, cannot capture all the complexities of real-world environments, leading to less precise localization results.

Non-geometric methods, which leverage pattern-matching algorithms or data-driven techniques, offer a promising alternative by not relying on predefined mathematical models of \ac{mpc} propagation. These approaches utilize environmental measurements and directly match prior observation-to-location databases using observed combinations of received signal metrics under the assumption of channel consistency. In \cite{tseng2017ray}, ray tracing is used during the offline phase to establish a prior CIR-to-location database. During the online phase, a K-Nearest Neighbors (KNN) algorithm is employed to match the measured \ac{cir} with entries in the database. \cite{gao2022toward} proposed a localization data generation method is proposed in \cite{gao2022toward} to aggregate \ac{mimo} information into a single image to achieve a more compact representation. Based on this compact representation, a multipath res-inception network is designed to perform position regression. A position regression method based on \ac{csi} is proposed in \cite{wang2016csi}, employing deep learning with a layer-by-layer training strategy during the training phase. Additionally, a semi-supervised approach for position regression is introduced in \cite{ruan2022ipos}, while the use of feature enhancement to improve recognition performance is highlighted in \cite{ruan2022hi}, such as manually extracting statistical features from the \ac{csi}. Non-geometric methods can handle complex and dynamic characteristics, offering more reliable and precise indoor positioning solutions compared to geometric methods. 

Several studies have explored position estimation in complex scenarios. In \cite{kram2019uwb}, the impact of complex reflection environments on \ac{cir} observations is qualitatively analyzed, and the feasibility of extracting complex environmental features from \ac{cir} measurements is discussed. The effects of various indoor propagation environments, including obstacles such as concrete walls, on \ac{cir} are examined in \cite{chen2020uwb}, along with testing the performance of several classical deep learning methods and genetic algorithms for position matching. The authors of \cite{witrisal2016high,leitinger2015evaluation} point out that the characteristics of diffuse reflectors cannot be described by any deterministic components. Previous studies, such as \cite{xu2023multi}, also mention the convergence challenges of estimated \ac{va}s associated with diffuse reflectors in geometry-based methods. To address this, a Bayesian particle-based method is proposed in \cite{wielandner2023multipath} to integrate the ambiguous \ac{va} generated by rough surfaces. Further exploration of diffuse reflectors' impact on the geometric estimation of specular reflection \ac{mpc}s is discussed in \cite{wilding2018accuracy}. In \cite{guo2024angle}, diffuse reflection is modeled by representing scattering points deterministically as an array of multiple scattering elements, approximating it as dense specular reflection, though this assumption is considered idealized. A similar modeling method, using tensor decomposition to estimate diffuse \ac{mpc}s within the reflective surface is presented in \cite{wen20205g}. The localization problem in the presence of dense \ac{mpc}s caused by low bandwidth is investigated in \cite{li2019massive}, where large \ac{mimo} antenna arrays are used to separately estimate the phase parameters of dense multipath and specular \ac{mpc}s, which are tightly coupled with \ac{toa}. These dense \ac{mpc}s are modeled as Gaussian noise with delay-dependent attenuation. Similarly, in \cite{ge20205g}, diffuse \ac{mpc}s are treated as dense specular \ac{mpc}s from rough surfaces, where effective diffuse \ac{mpc}s were considered specular \ac{mpc}s caused by random points on the rough surface. The distribution of these random points was determined by the surface's roughness. This work used a clustering approach to define diffuse reflectors, aiming to estimate and cluster the effective \ac{va}s of diffuse reflectors. These diffuse reflection mapping methods all assume or require that \ac{mpc}s are distinguishable.

This work focuses on a complex multipath environment where specular reflection and diffuse reflection coexist. Distinct from the aforementioned research, our work models diffuse reflectors from the perspective of uncertainty, analyzing the statistical distribution of power-delay characteristics of \ac{mpc}s reflected by diffuse and specular surfaces. Based on this theoretical foundation, we propose a variational inference-based semi-supervised deep learning method for filtering localization information. By using \ac{cir} as input, the method isolates indistinguishable \ac{mpc}s that do not contribute to localization, such as diffuse \ac{mpc}s, achieving both compression of localization information and improvement in positioning accuracy.





The contributions of this work are summarized as follows.
\begin{itemize}
    \item We modeled the physically existing diffuse reflectors from the perspective of uncertainty, demonstrating that the aliasing \ac{mpc}s caused by diffuse reflectors and those caused by specular reflectors are independent and separable.
    \item We proposed a variational inference-based semi-supervised deep learning framework—Multipath Disentanglement Network (MudiNet). This model leverages task-guided latent space to disentangle \ac{mpc}s that do not contribute to positioning, thereby enhancing positioning performance.
    \item We utilized the 5G signaling-controlled transmission power adjustment mechanism to simulate multipath aliasing using different SNR conditions. Simulation results indicate that our method achieves comparable accuracy under high-quality signal conditions while exhibiting greater robustness in scenarios with severe multipath aliasing.
\end{itemize}

The structure of this paper is organized as follows. Section II formulates the positioning problem. The signal model and diffuse-specular reflection disentanglable channel model are also provided in Section II. Section III details the proposed positioning method. Simulation results are presented in Section IV, followed by the conclusion of this work in Section V.

\emph{Notations and Symbols:} Scalars are denoted by lowercase italic letters (e.g., $a$). Vectors are represented by lowercase boldface letters (e.g., $\mathbf{a}$), while matrices are denoted by uppercase boldface letters (e.g., $\mathbf{A}$). $(\cdot)^\top$, $(\cdot)^{-1}$, $\text{tr}(\cdot)$, $\text{vec}(\cdot)$ and $\| \cdot \|_2$ indicate the transpose, inverse, trace, vectorization, and $\ell$-2 norm operations, respectively. $(\cdot)_{i }$ represents the $i$-th element of the vector. $(\cdot)_{i \cdot}$, $(\cdot)_{\cdot j}$, and $(\cdot)_{i j}$ represent the $i$-th row, the $j$-th column, and the element in the $i$-th row, $j$-th column of a matrix, respectively. $(\cdot)_{i:j,m:n}$ is the submatrix formed by the rows from $i$ to $j$ and the columns from $m$ to $n$ of a matrix. 

\section{Problem formulation}
In this section, we will introduce the \ac{map} under complex reflection scenario, and a \ac{dsrd} channel model.
\subsection{Multipath-assisted Positioning under Specular and Diffuse Coexist Situation}
\begin{figure}
    \centering
    \includegraphics[width=1\linewidth]{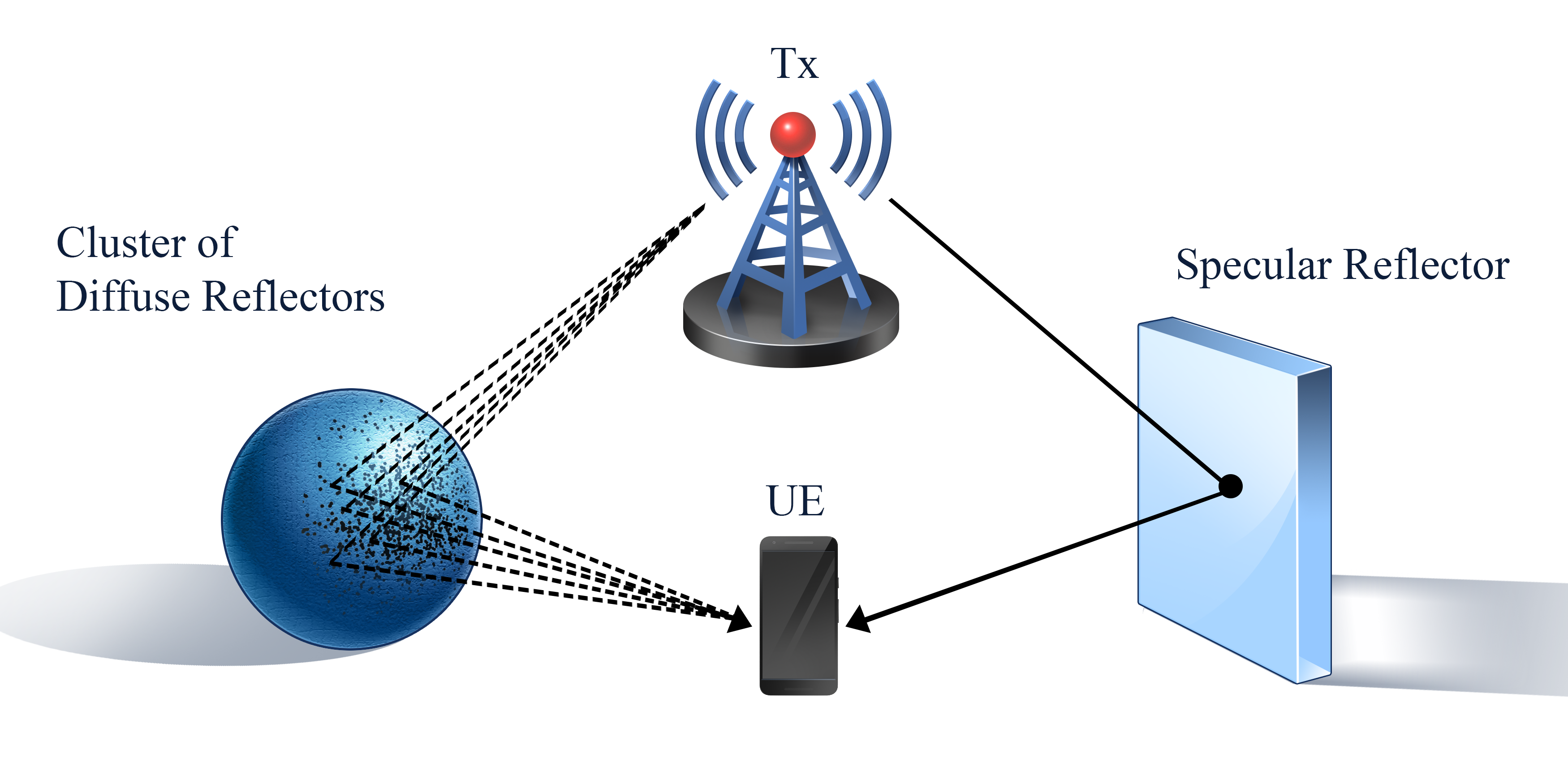}
    \caption{The left side represents the diffuse \ac{mpc}s, where the points of interaction between the Tx and UE on the reflector are considered to be randomly distributed within a certain region. The right side represents the specular \ac{mpc}, where the points of interaction between the Tx and UE on the reflector are uniquely determined.}
    \label{fig:reflector}
\end{figure}
Consider a typical \ac{wss} wireless environment with a transmitter (Tx) and a \ac{ue} receiver. The signal transmitted from the Tx reaches the \ac{ue} after being reflected by various reflectors, so the received signal at the \ac{ue} can be represented as a superposition of reflected \ac{mpc}s with different time delays. Equation \ref{distribution} represents the \ac{mpc} as an ideal model, each \ac{mpc} corresponds to a \ac{va}. However, the presence of diffuse reflection introduces additional randomness to the \ac{mpc}s. The \ac{mpc}s caused by scatterers arises from the diffusion of signal energy in various directions when electromagnetic waves strike rough scatterers. In modeling, existing studies often treat all reflectors including both scatterers (diffuse reflectors) and specular reflectors in the environment as a unified random distribution throughout the entire space \cite{jiang20193,cheng2022channel,bai20213}. However, in reality, scatterers are fixed physical entities, such as indoor plants or rough walls. This study differentiates between diffuse reflectors and specular reflectors in the spatial model, treating the specular reflection components as deterministic multipath, while considering the diffuse reflection components as uncertain, indistinguishable multipath, as shown in Figure \ref{fig:reflector}. For different types of reflectors, the signal model can be formulated as a combination of two components: specular component $CIR_{s}(t,\tau)$ and diffuse component $CIR_{d}(t,\tau)$.
\begin{equation}
    \label{cir}
    CIR(t,\tau) = CIR_{s}(t,\tau) + CIR_{d}(t,\tau) + \textbf{n}(t)
\end{equation}
Specifically, the two separated parts of the $CIR(t,\tau)$ can be expressed as:
\begin{equation}
    \label{distribution}
    \begin{split}
    CIR_{s}(t,\tau)&=\sum^{N_s}_{i=0}a_i(t)\delta(t-\tau_i) \xrightarrow[]{}(\tau_s,\beta_s) \\
    CIR_{d}(t,\tau)&=N_d\int_{D}f_db_j(t)S(t-\tau_j) d\tau\xrightarrow[]{}(\boldsymbol{\tau}_d,\boldsymbol{\beta}_d)
    \end{split}
\end{equation}


Here, the term $a_i(t)$ denotes the time-varying power of the $i_{th}$ specular \ac{mpc}s, where $i=0$ indicates the \ac{fp}, and $N_s$ represents the total number of specular \ac{mpc}s, including the \ac{fp}. The term $\textbf{n}(t)$ represents the noise component. For the specular reflection component, the interaction points on the reflective surface between the Tx and UE are uniquely determined; they cannot be modeled as a random distribution, but only as a \ac{mpc} set. In the localization task, the key multipath set parameters of interest are the delay set $\tau_s$ and the power set $\beta_s$.

Similarly, $b_j(t)$ represents the time-varying power of the $j_{th}$ diffuse \ac{mpc}s, with $N_d$ denoting the number of observed diffuse \ac{mpc}s. Considering that the parameters of interest in positioning tasks are typically delay distribution $\boldsymbol{\tau}_d$ and power distribution $\boldsymbol{\beta}_d$ of diffuse \ac{mpc}s, we attempt to establish a statistical model for localization parameters under diffuse reflection. Assuming that the diffuse reflectors in region $D$ follow a Gaussian distribution $f_d$ and that the \ac{ue}'s observation of the diffuse reflectors follows a uniform distribution, we introduce the observation function:
\begin{equation}
    S(t-\tau_j)=
    \begin{cases}
    \delta(t-\tau_j) & \text{diffuse reflector \textit{j} is observed}\\ 
    0 & \text{diffuse reflector \textit{j} is not observed}
    \end{cases}
\end{equation}

In a positioning environment where diffuse and specular reflections coexist, our objective is to determine the conditional distribution from the \ac{cir} to the estimated 2D position of the \ac{ue}:
\begin{equation}
\label{positioning}
    \hat{\textbf{p}}=(\hat{\mathbf{p}}_x,\hat{\mathbf{p}}_y)=p(\hat{\textbf{p}}|\boldsymbol{\tau}_d,\boldsymbol{\beta}_d,\tau_s,\beta_s)
\end{equation}
In this study, we use two single anchor \ac{siso} positioning schemes as case studies to emphasize the role of \ac{mpc}s. In the first scheme, we introduce a specular reflective wall commonly found in typical environments. In the second scheme, we incorporate both specular reflection points and diffuse scatterers, such as indoor plants, to simulate more complex reflection dynamics.

\subsection{DSRD Channel Model}
The contributions of \ac{mpc}s caused by diffuse reflectors and specular reflectors to positioning are different. For specular reflection, as the \ac{ue} moves, the reflection points can be determined by the mirror image of the reflective surface, resulting in a fixed \ac{va} for each position. 
If we assume that the specular reflection point of the transmitted signal on the surface is $(x_s,y_s)$, then at time $t_0$,the corresponding delay is given by:
\begin{equation}
    \tau_s(t_0) = \sqrt{x_s^2+y_s^2} + \sqrt{(\mathbf{p}_x(t_0)-x_s)^2+(\mathbf{p}_y(t_0)-y_s)^2}
\end{equation}
Correspondingly, the received power can be expressed as:
\begin{equation}
    \beta_s(t_0) = L^2\frac{P}{r_sr_t}
\end{equation}
where $r_s$ denotes the distance between the Tx and specular reflection point(incident path), $r_t$ denotes the reflection path and $L$ the power attenuation coefficient, representing the ratio of propagation distance to power attenuation in free space. As shown above, if observation noise is not considered, the specular reflection paths are deterministically modeled without randomness. In the complete received CIR, all randomness of the specular reflection multipath components originates from the generalized stationary Gaussian noise $\mathbf{n}(t)$.

\begin{figure}
    \centering
    \includegraphics[width=1\linewidth]{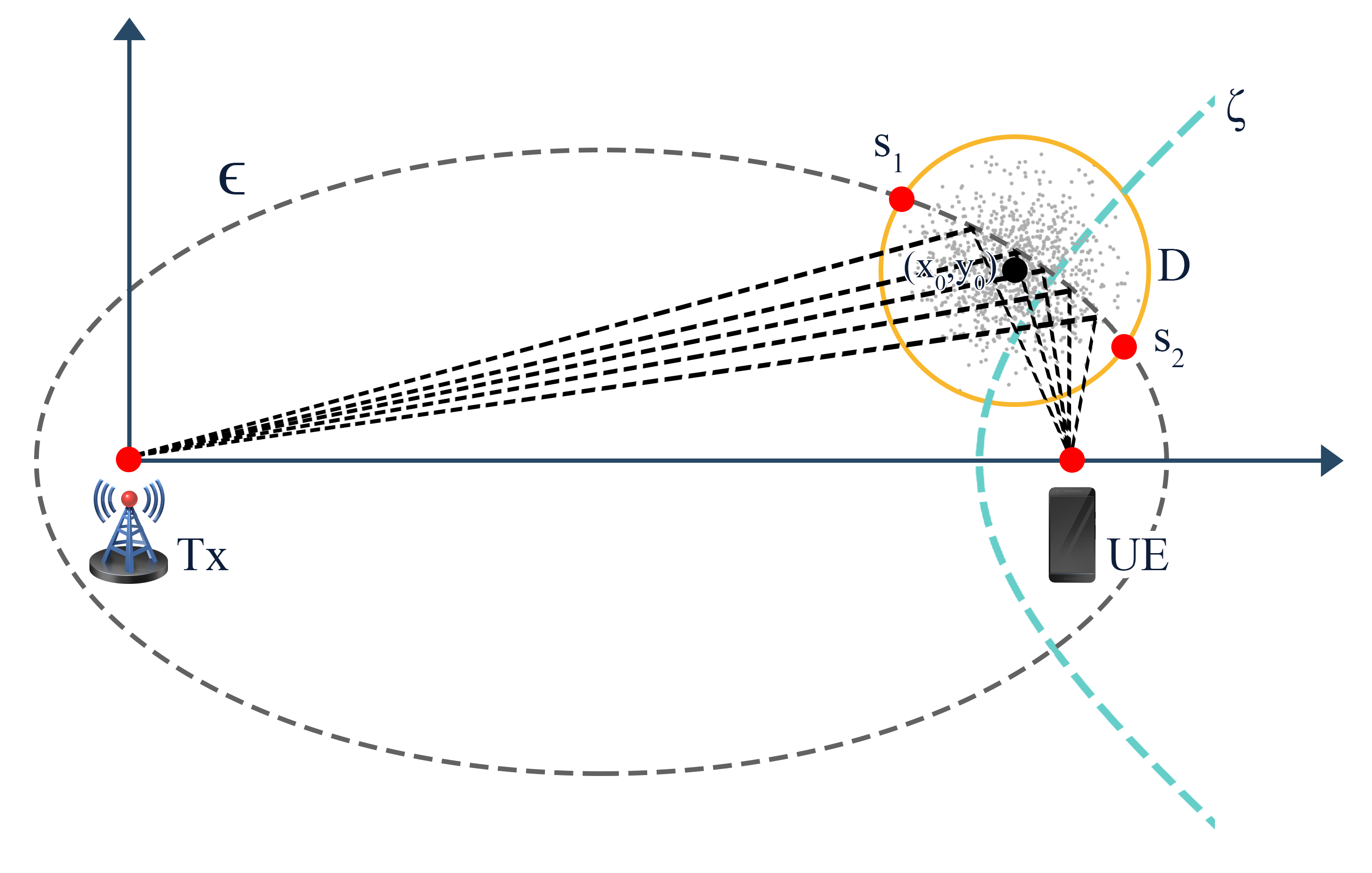}
    \caption{The illustration of diffuse reflector distribution. The signal is transmitted from Tx, reflected by the scatterer region $D$, and received at UE, with an incident path length of $r_d$ and a reflected path length of $r_t$.}
    \label{fig:diffuse}
\end{figure}

In contrast, for diffuse reflection, the scatter points change as the \ac{ue} moves, leading to variations in the reflect points. Therefore, using the \ac{mpc} caused by diffuse reflection for positioning typically results in difficulty in convergence of the positioning results\cite{xu2023multi}. In this study, as shown in Figure \ref{fig:diffuse} scatterers are considered to be locally Gaussian distributed $f_d(x,y)$ around a central position $(x_0,y_0)$:
\begin{equation}
f_d(x,y)=\frac{1}{2\pi\sigma_x\sigma_y}\exp{(-\frac{(x-x_0)^2}{2\sigma_x^2}-\frac{(y-y_0)^2}{2\sigma_y^2})},x,y\in D
\end{equation}
where $\sigma_x$, $\sigma_y$ represent the standard deviations of the scatterer distribution, which are taken as the major and minor axes of the scatterer distribution region $D$. 

\textit{Proposition 1}: Given a position of \ac{ue}, $CIR_s(t,\tau)$ is conditionally independent with $CIR_d(t,\tau)$.

~\textit{Proof}:
~Considering a single scatterer, there exists a subregion within $D$ that shares the same delay observation. The probability density of delay $\tau_d$ caused by scattering in the \ac{cir} is:
\begin{equation}
    \boldsymbol{\tau}_d(t_0)=f(t_0,\tau)=\int^{s_{2}(t_0,\tau)}_{s_{1}(t_0,\tau)}f_d(x,y)ds
\end{equation}
As shown in Figure, giving a \ac{ue} position at time step $t=t_0$, $\epsilon$ represents an ellipse defined with $\textbf{p}(t_0)$ and $\textbf{p}_0$ which correspond to the position of \ac{ue} and Tx as foci, $s_1$ and $s_2$ are the endpoints of the arc segment where $\epsilon$ intersects region $D$. The \ac{mpc}s generated along the arc segment $s_{1,2}(t_0,\tau)$ have the same delay.

For the power distribution $\beta_d$ of this scatterer, we first consider the incident power on the scatterer:
\begin{equation}
    P_{d} = L\frac{P}{r_d}
\end{equation}
let $P$ be the transmitted power from the Tx, $P_d$ the power received at the scatter point, $r_d$ the distance between the scatterer and the Tx antenna (incident path). Assuming the power reflected by the scatterer is proportional to the distribution probability $f(x,y)$ of the scatterer in the differential area element $dS$ within region $D$, we consider the ellipse $\epsilon=r_t+r_r$ and a confocal hyperbola $\zeta=r_r-r_t$ (where the incident path $r_t$ is assumed to be less than the reflected path $r_r$). At the intersection points, $\epsilon$ and $\zeta$ are orthogonal, forming a differential area element $dS=d\epsilon d\zeta$. Therefore, the contribution of a differential surface element $dS$ to the received power $dP_r$ is:
\begin{equation}
    dP_r=L\frac{P_df_d(x,y)dS}{r_r}=L^2\frac{Pf_d(x,y)}{r_dr_t}d\epsilon d\zeta
\end{equation}
Thus, the distribution of scattered power in the delay domain should be represented as the synthesis of scattered power along multiple intersecting paths $s_{1,2}(\tau)$. This involves integrating over all surface elements along the intersecting curve $s_{1,2}(\tau)$:
\begin{equation}
    \boldsymbol{\beta}_d(t_0)=P_r(t_0,\tau)=\int^{s_2(t_0,\tau)}_{s_1(t_0,\tau)}L^2P\frac{f_d(x,y)}{r_tr_d}|_{(x,y)=(\epsilon,\zeta)}dxdy
\end{equation}
In summary, given a \ac{ue}'s position, the \ac{mpc}s from diffuse reflectors depend on the spatial distribution of the scatterers and the observation noise, whereas for specular reflectors, the randomness in the \ac{cir} arises solely from the observation noise $\textbf{n}(t)$. Under the assumption of WSS, given a \ac{ue} position at time step $t_0$, the $CIR_d(t,\tau)$ from diffuse reflection is conditionally independent of the $CIR_s(t,\tau)$ from specular reflection:
\begin{equation}
\label{ci}
\begin{split}
        p((\boldsymbol{\tau}_d,\boldsymbol{\beta}_d),(\tau_s,\beta_s))|\textbf{p}(t_0)) = \\
      p((\boldsymbol{\tau}_d,\boldsymbol{\beta}_d)|\textbf{p}(t_0))p((\tau_s,\beta_s)|\textbf{p}(t_0))    
\end{split}
\end{equation}
Therefore, we consider that in a spatial channel where specular reflectors and diffuse reflectors coexist, the \ac{mpc}s received are separable when observing the specular and diffuse reflectors with a known \ac{ue} position, which makes \textit{Proposition 1} proved.



\section{Proposed Method}

\begin{figure}
    \centering
    \includegraphics[width=1\linewidth]{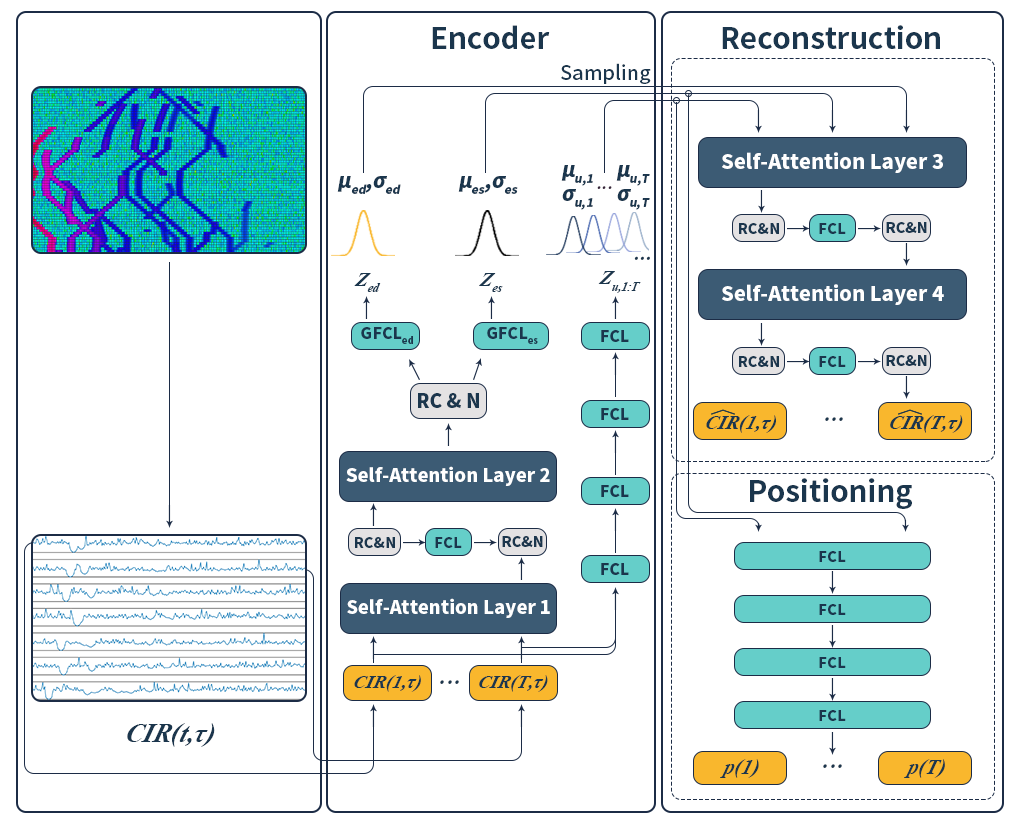}
    \caption{Structure of Proposed Multipath Disentangled Network(MudiNet), RC denotes the residual connection and layer normalization, FCL means fully connected layer. GFCL means the input the means of multi-time self-attention output to FCL, to obtain the global estimation of environment.}
    \label{fig:overview}
\end{figure}

In this section,  we will introduce the variational inference based latent variable modeling, a \ac{dsrd} neural network, and a MAP network based on diffuse reflections suppression. The network structure is presented in Figure\ref{fig:overview}

\subsection{Latent Variable Model of Multi-time CIRs}
The \ac{lvm} is a type of statistical model that introduces unobserved latent variables to capture the underlying structure or complex relationships within the data. Numerous studies \cite{zhang2019learning,cifka2021self,liu2021multiple} have demonstrated that \ac{lvm}s can facilitate the disentangling of the latent structure in data. 

In general multipath channel environments, the \ac{cir} features can typically be decomposed into two groups: \ac{ue} position-related features and environmental features. Previous works have extensively modeled these two components mathematically \cite{xu2023multi,gentner2016multipath,leitinger2019belief,amiri2023indoor}. In these models, \ac{ue} position-related features include information such as ranging and angle measurements, while environmental features encompass aspects such as the distribution and material properties of reflectors. Moreover, it has been demonstrated in related research that specular environmental features and \ac{ue} position-related features are independent \cite{wang2024multipath,li2023variational}.

From Equation (\ref{ci}), it follows that the \ac{cir} components generated by diffuse scatterers and those generated by specular reflectors are conditionally independent given the \ac{ue}'s position. However, the observation of diffuse scatterers from a single \ac{cir} at a specific time is incomplete. The cutting arc $s(t_0,\tau)$ formed by a single focal ellipse struggles to capture the distribution of diffuse reflectors across the entire region $D$, and the observation results are highly dependent on the \ac{ue}'s position at that moment. Therefore, as illustrated in the figure, we aim to use multi-time \ac{cir} observations to enhance the observation of diffuse scatterers, to facilitate the disentangling of diffuse environmental features from \ac{ue} position related-features.

Building upon the analysis of the independence of diffuse \ac{cir} component and specular component \ac{cir}, these findings motivate us to use \ac{lvm}s to further disentangle the feature structure of the \ac{cir}, thereby mitigating the adverse impact of diffuse reflectors on positioning. In this study, we define three sets of latent variables: the environment-related variable $\textbf{z}_d$ associated with diffuse scatterers, the environment-related variable $\textbf{z}_s$ associated with specular reflectors, and the UE position-related variable $\textbf{z}_u$.

\textit{Proposition 2}: Given a multi-time \ac{cir} observation sequence, if the reflectors are sufficiently observed within this sequence by \ac{ue}, then the \ac{lvm} $\textbf{z}_d$ and $\textbf{z}_u$ are independent.

~\textit{Proof}:
From Equation \ref{ci}, it can be deduced that in a single observation, the \ac{ue}'s observations of diffuse scatterers and specular reflectors are conditionally independent given the \ac{ue}'s position. This is because, when the \ac{ue}'s position is fixed, the elliptical curve's intersection with region $D$, denoted as the arc $s_{1,2}(t_0,\tau)$, is unique. Hence, the arc $s$ can be expressed as a function of time. As the \ac{ue} performs redundant observations of the diffuse reflectors, the set of arcs approaches region $D$:
\begin{equation}
    \{s_{1,2}(1:T,\tau)\}\approx D
\end{equation}
Thus, the delay-power distribution observation from diffuse reflector region $D$ becomes disentangled from time and \ac{ue}'s position.
\begin{equation}
\begin{split}
    \boldsymbol{\tau}_d(t=1:T)&=\int_{D}f(x,y)dxdy \\
    \boldsymbol{\beta}_d(t=1:T)&=L^2P\int_{D}\frac{f(x,y)}{r_tr_r}d\epsilon d\zeta
\end{split}
\end{equation}
Here, $T$ represents the total number of time steps in the \ac{ue} trajectory. Under joint observations of the \ac{cir}s at multiple time steps, the observation of region $D$ becomes independent of time. Consequently, $\textbf{z}_d$ and $\textbf{z}_u$ are independent, 
\begin{equation}
    \begin{split}
        p(\textbf{z}_d,\textbf{z}_u)=p(\textbf{z}_d)p(\textbf{z}_u)
    \end{split}
\end{equation}
which makes \textit{Proposition 2} proved. Additionally, from Equation \ref{ci}, it can be inferred that $\textbf{z}_s$ and $\textbf{z}_d$ are independent. Based on these conclusions, we can further establish the relationship between input variables, latent variables, and the parameters to be estimated.

\subsection{Self-Attention based Feature Extraction of Multi-time CIRs}
We define the logarithm of multi-time CIR observations as the input variable $\textbf{x}$ to the model, which encapsulates the delay and power information from different reflectors: $\textbf{x}=\{\textit{x}_i,i=0:T\}$. 

\begin{figure}
    \centering
    \includegraphics[width=1\linewidth]{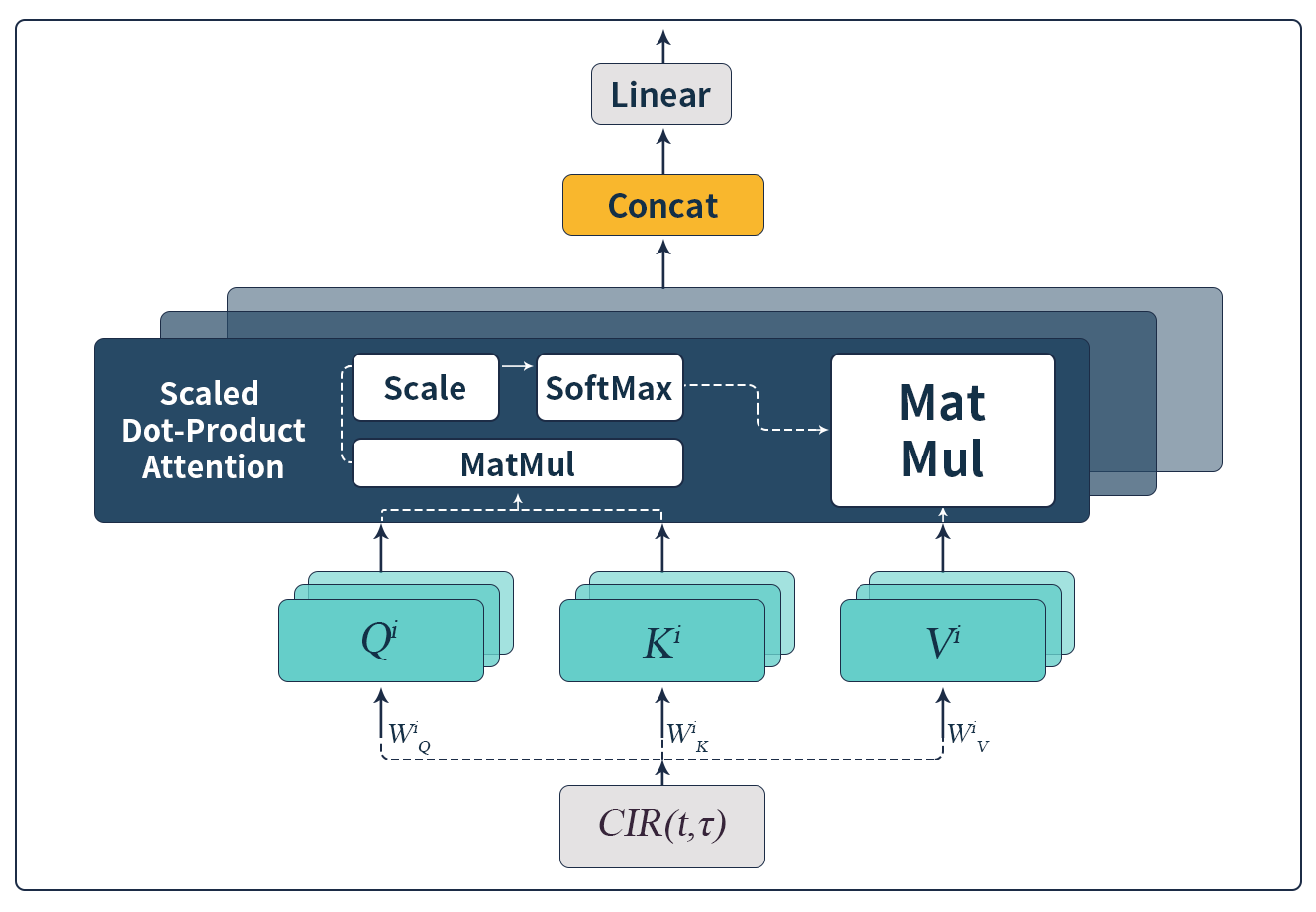}
    \caption{Structure of Self-attention Layer}
    \label{fig:self-attention}
\end{figure}

In prior studies, the joint observation of multi-time CIRs has been demonstrated as an effective approach for environmental estimation. However, traditional methods such as Kalman filtering fail to reliably estimate the distribution of diffuse reflectors\cite{xu2023multi}. Here, we aim to leverage neural networks to effectively integrate the observational information of diffuse reflectors, facilitating subsequent latent variable estimation and disentanglement. The self-attention mechanism\cite{vaswani2017attention}, recognized as a suitable structure for extracting temporal features, enables the encoder to integrate information across multi-time CIRs. It recalibrates weights to CIR observations at different times, thereby achieving a weighted estimation of environmental variables. Specifically, as shown in Fig. \ref{fig:self-attention} we apply self-attention along the temporal dimension to the input data $\textbf{x}$:
\begin{equation}
    \begin{split}
    Q_i=\textbf{x}_iW_{Q_i}\\
    K_i=\textbf{x}_iW_{K_i}\\
    V_i=\textbf{x}_iW_{V_i}        
    \end{split}
\end{equation}

The self-attention structure projects $\textbf{x}$ at each time step into an attention space $\mathbb{R}^d$ and computes attention scores for every pair of $\textbf{r}$ at different time steps:
\begin{equation}
    Score_{i,j}=\frac{Q_iK_j^\top}{\sqrt{d}}
\end{equation}
where $d$ is the dimension of the attention space. The attention scores quantify the contribution of \ac{cir} observations at different time steps to the estimation of environmental variables. These scores are further approximated into conditional probabilities via the Softmax function:
\begin{equation}
    p(K_j|Q_i)\overset{\sim}{=}Softmax(Score_{i,j})=\frac{\exp{Score_i,j}}{\sum^T_{j=1}\exp{Score_i,j}}
\end{equation}
The output of the self-attention mechanism $\textbf{r}=\{\textbf{r}_i,i=0:T\}$ represents the weighted expectation of temporal \ac{cir}s:
\begin{equation}
    \textbf{r}_i=\mathbb{E}(V_j|Q_i)=\sum^T_{j=1}p(K_j|Q_i)V_j
\end{equation}

\subsection{Variational Inference Loss Function and Evidence Lower Bound}

Given a set of input data $\textbf{x}$ and a feature $\textbf{r}$, there exists a conditional distribution $p(\textbf{z}_s,\textbf{z}_d|\textbf{r})$ and $p(\textbf{z}_u|\textbf{x})$, which facilitates the inference of the feature structure within $\textbf{x}$. These two conditional distributions form the encoder.
\begin{equation}
    Enc(\textbf{x},\textbf{r})\sim\{p(\textbf{z}_s,\textbf{z}_d|\textbf{r}),p(\textbf{z}_u|\textbf{x})\}
\end{equation}

Variational inference requires knowledge of the variables and their distributions. Since the conditional distribution $p(\textbf{z}_s,\textbf{z}_d|\textbf{r})$ and $p(\textbf{z}_u|\textbf{x})$ are unknown, we introduce variational distributions $q(\textbf{z}_d|\textbf{r})$, $q(\textbf{z}_s|\textbf{r})$ and $q(\textbf{z}_u|\textbf{x})$ to approximate the intractable true posterior. Under the self-attention architecture, the estimation of the environmental variables becomes a weighted \ac{mle} based on self-attention: 
\begin{equation}
    L(\textbf{z}_s,\textbf{z}_d|\textbf{r})=\prod^T_{i=1}p(\textbf{r}_i|\textbf{z}_s,\textbf{z}_d)
\end{equation}
The information related to specular reflection is considered contributive to positioning results, while the information related to diffuse scattering is not. Therefore, assuming $\textbf{z}_s$ and $\textbf{z}_d$ are effectively decoupled, position estimation based on the input variables can be equivalently reduced to position estimation based on the latent variables $\textbf{z}_s$ and $\textbf{z}_u$.
\begin{equation}
\label{pos}
    p(\hat{\textbf{p}}|\textbf{x}) \sim p(\hat{\textbf{p}}|\textbf{z}_u,\textbf{z}_s)=f_{pos}(\textbf{z}_u,\textbf{z}_s)
\end{equation}
Variational inference is employed to estimate and disentangle the latent variables embedded in $\textbf{x}$, achieved through the minimization of the \ac{elbo}:
\begin{equation}
\label{elbo}
\begin{split}
        ELBO(q;\textbf{r};\textbf{x})=\mathbb{L}_{rec}-\\KL(q(\textbf{z}_d|\textbf{r})||p(\textbf{z}_d))-\\KL(q(\textbf{z}_s|\textbf{r})||p(\textbf{z}_s))-\\KL(q(\textbf{z}_u|\textbf{x})||p(\textbf{z}_u))
\end{split}
\end{equation}
The \ac{elbo} consists of two components: the Kullback-Leibler (KL) divergence constraint, which enforces the disentanglement of latent variables, and the reconstruction error constraint, derived from the symmetrical reconstruction structure:
\begin{equation}
\label{dec}
    Rec(\textbf{z}^{(r)}_u,\textbf{z}^{(r)}_d,\textbf{z}^{(r)}_s)\sim p(\textbf{x}|\textbf{z}^{(r)}_u,\textbf{z}^{(r)}_d,\textbf{z}^{(r)}_s)
\end{equation}
where $KL(\cdot||\cdot)$ denotes the KL divergence between two distributions. $\mathbb{L}_{dec}(\textbf{x},\hat{\textbf{x}})$ denotes the reconstruction loss between the input \ac{cir}s $\textbf{x}$ and reconstructed \ac{cir}s $\hat{\textbf{x}}$, we employ \ac{mse} loss as the reconstruction loss: 
\begin{equation}
    \mathbb{L}_{rec}=\mathbb{E}_{q(\textbf{z}_s,\textbf{z}_d,\textbf{z}_u|\textbf{x})}log(p(\textbf{x}|\textbf{z}_s,\textbf{z}_d,\textbf{z}_u))
\end{equation}

Environmental features require the integration of global information, which in this context is estimated using the temporally weighted features $\textbf{r}$. In the absence of a \ac{ue} motion constraint, \ac{ue}'s position information is considered to be independent of temporal sequences. Therefore, instead of using $\textbf{r}$, the raw input $\textbf{x}$ is employed to estimate temporal features. This approach is expected to mitigate the introduction of incorrect information in trajectory estimation.

\subsection{Semi-supervised Algorithm with Position Labels}

To train the model to disentangle latent variables and estimate positions, it is essential to minimize the ELBO defined in Equation (\ref{elbo}). Following the recommendations in \cite{kingma2013auto}, we assume prior distributions for the latent variables, which are typically modeled as Gaussian distributions:
\begin{equation}
    \begin{split}
        \textbf{z}_s\sim N(0,\textbf{I}\epsilon_s^2)\\
        \textbf{z}_d\sim N(0,\textbf{I}\epsilon_d^2)\\
        \textbf{z}_u\sim N(0,\textbf{I}\epsilon_u^2)\\
    \end{split}
\end{equation}
Here, $\textbf{I}\in \mathbb{R}^l$ denotes a unit vector with dimensionality $l$ in the latent space $\mathbb{R}^l$, $\textbf{z}_s,\textbf{z}_d$ and $\textbf{z}_u$ are typically assumed to follow a standard Gaussian distribution. However, to facilitate the disentanglement of different latent variables, we define their prior uncertainties following the principle: $\epsilon_u^2>\epsilon_d^2>\epsilon_s^2$. This choice is motivated by the preceding analysis: the uncertainty of specular reflections depends solely on observation noise, while diffuse reflections are influenced by both observation noise and the distribution of diffuse reflectors. Moreover, in random motion, position-related variables are expected to exhibit the highest level of uncertainty.

The latent variables $\textbf{z}_s$, $\textbf{z}_d$, and $\textbf{z}_u$, represented by probability distributions, pose challenges for gradient propagation within the network. To address this, we follow the suggestion in \cite{kingma2013auto} and employ the reparameterization trick during the reconstruction and subsequent positioning stage, in Equation (\ref{dec}) transforming the probability distributions into continuous feature vectors through resampling:
\begin{equation}
    \textbf{z}^{(r)}=\hat{\mu}+\hat{\sigma}^2\epsilon^2\\
\end{equation}
where $\hat{\mu}$ and $\hat{\sigma}^2$ represent the mean and variance of the latent variables as estimated by the $Enc$.
In the previous sections, we outlined the methods for feature extraction and latent variable disentanglement. However, disentanglement alone does not ensure that the latent variables encode the desired information. To guide the representation of information in the latent space and simultaneously accomplish the subsequent positioning task, we adopt a semi-supervised learning algorithm. 

This process is supported by incorporating position labels into the dataset. Given a sample space $\mathbb{R}^N$ with the label of a sample $\textbf{p}^{(j)}=\{\textbf{p}_i^{(j)},i=0:T,j=1:N\}$, we define the positioning loss using the \ac{mse} based on Equation(\ref{pos}):
\begin{equation}
\label{eq:mse}
    \mathbb{L}_{pos}=\frac{1}{N}\sum_{j=1}^N\|f_{pos}(\textbf{z}_u^{(r)},\textbf{z}_s^{(r)})-\textbf{p}^{(j)}\|_2
\end{equation}
Thus, the overall loss function $\mathbb{L}$ consists of both the \ac{elbo} and the positioning loss:
\begin{equation}
    \mathbb{L}= ELBO(q;\textbf{r};\textbf{x})+\mathbb{L}_{pos}
\end{equation}
During the training phase, we employ the power delay profiles of multi-time \ac{cir}s with position labels as input and optimize the overall loss function $\mathbb{L}$ to learn the parameters of encoder $Enc$ function, reconstruction $Rec$ function, and positioning model $f_{pos}$ function. During the testing phase, all model parameters are frozen, and the input test \ac{cir}s are directly fed into the trained network structure for position estimation.

\section{Simulation Results}
\subsection{Simulation Setting and Dataset Generation}
\begin{figure}[h]
    \centering
    {\includegraphics[width=0.8\linewidth]{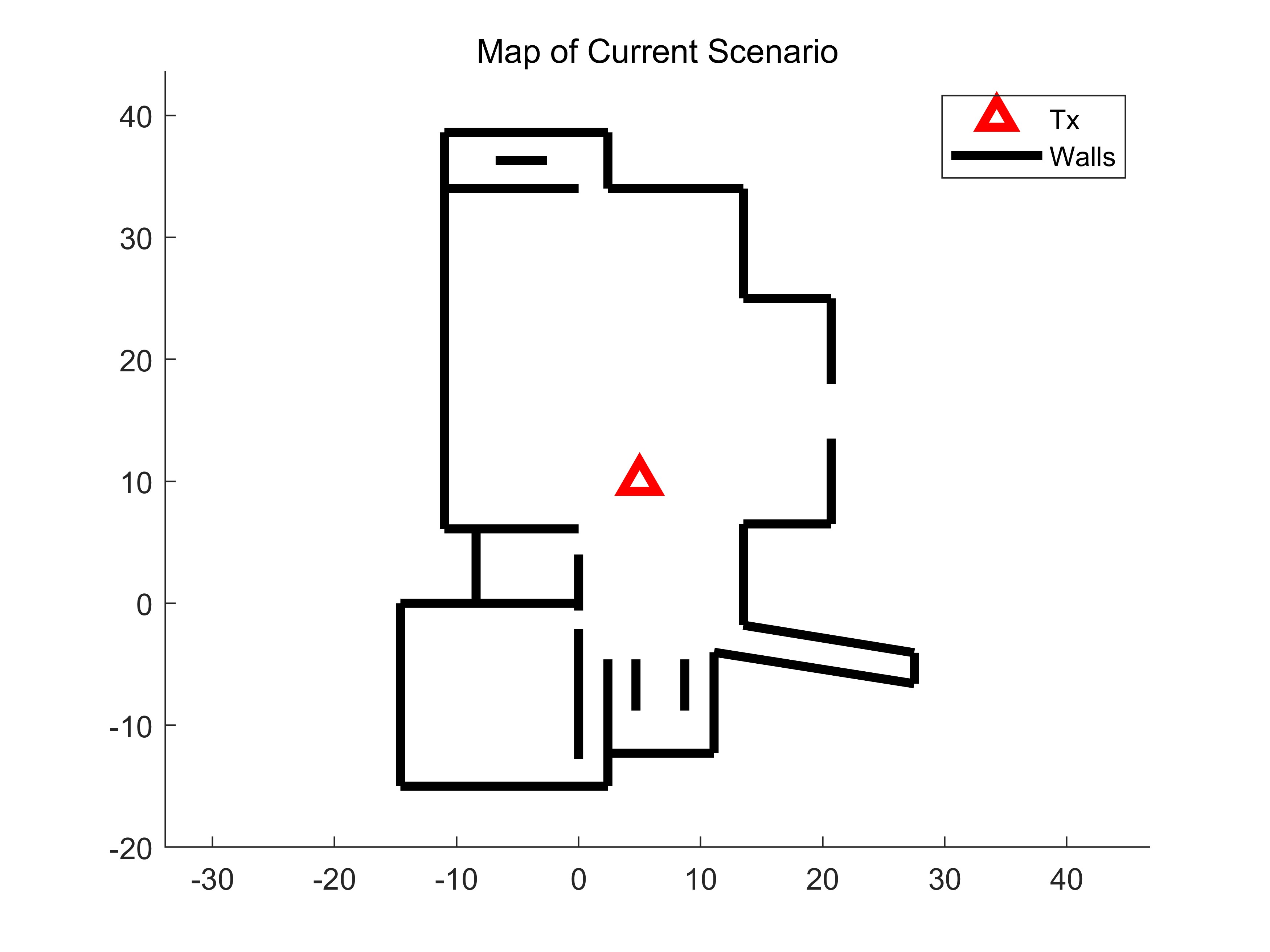}}
    
    \caption{Simulation Scenarios Illustration. Representing Scenario, where the walls are defined as smooth ideal reflective surfaces.}
    \label{fig:scene}
\end{figure}

We use the MATLAB R2024b simulation platform to simulate an indoor wireless environment under different degrees of multipath aliasing conditions. and analyze the positioning performance of the proposed method under different multipath aliasing conditions. All experiments are implemented into one scenario. As shown in Figure \ref{fig:scene}(a), the scenario represents a classical environment where reflectors consist solely of ideal smooth reflective surfaces. The scenario is generated using a real indoor map.
We design the simulation following the standard 5G channel parameters, the basic wireless parameters for the simulation are summarized in Table \ref{tab:params}. We use the simplified Friis transmission equation to simulate the process of signal free-space attenuation, as shown in Equation (\ref{friis}):

\begin{equation}
\label{friis}
    Rss = L\times(\frac{c}{4\pi \lambda})^2/(dist^2)
\end{equation}
where $Rss$ represents the received signal power without noise, $\lambda$ denotes the wavelength, $dist$ is the distance between the \ac{ue} and the Tx (also Reflector). The Reflection Factor denotes the power attenuation factor $L$ for each signal reflection. In this simulation, the maximum order of reflection is set to 2.
\begin{table}[h]
    \centering
    \renewcommand\arraystretch{1.2}
    \caption{The Basic Parameter Configuration of Simulation}
    \begin{tabular}{cc}
    \toprule
        \textbf{Parameter} & \textbf{Configuration} \\
    \hline
        Bandwidth & 100 MHz\\
        
        Sampling Factor & 4 \\
    
        Reflection Factor/ L & 0.25\\
    
        Radio Frequency/ $\frac{1}{\lambda}$ & 2.4 GHz\\

        Speed of Light/ c & $3\times10^8$ m/s\\

        Transmit Power & 1 \\

        Background Noise & $1\times10^{-10}$ \\

        Spread Gain/ $G_{ss}$ & 20 $dB$\\

        Trajectory Length / s & 110 \\

        Maximum Number of Taps & 300 \\
        
    \bottomrule
    \end{tabular}
    \label{tab:params}
\end{table}

We construct the dataset by randomly generating trajectories and their corresponding CIR observation sequences on the two scenarios. The trajectory generation process starts with a randomly chosen starting position and orientation, followed by trajectory generation at a fixed velocity of $1 m/s$ and a randomly selected heading angle from the interval $[-\pi/2,\pi/2]$. The total length of the trajectory is controlled to 110 seconds. If the \ac{ue} contacts a wall and is predicted to pass through it in the next second, the trajectory generation is stopped, and the trajectory is discarded. For each experiment, 5000 trajectory-CIR observation pairs are independently generated, with 80\% used as the training set and the remaining 20\% used as the test set.

\begin{figure*}[htbp]
    \centering
    \begin{minipage}{\textwidth}
        \centering
        \begin{minipage}{0.26\textwidth}
            \centering
            \includegraphics[width=\textwidth]{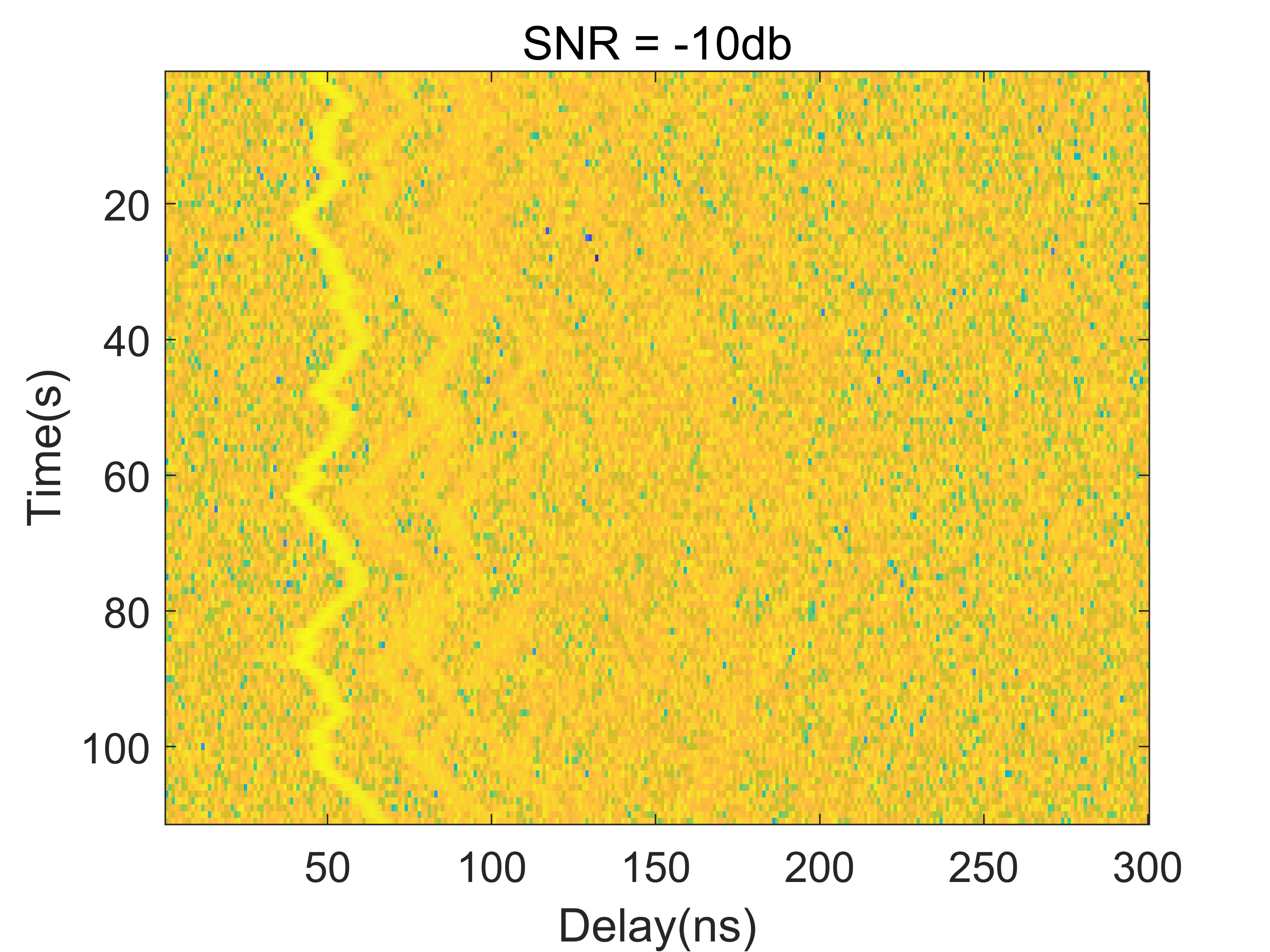}
        \end{minipage}
        \hspace{-0.5cm}
        \begin{minipage}{0.26\textwidth}
            \centering
            \includegraphics[width=\textwidth]{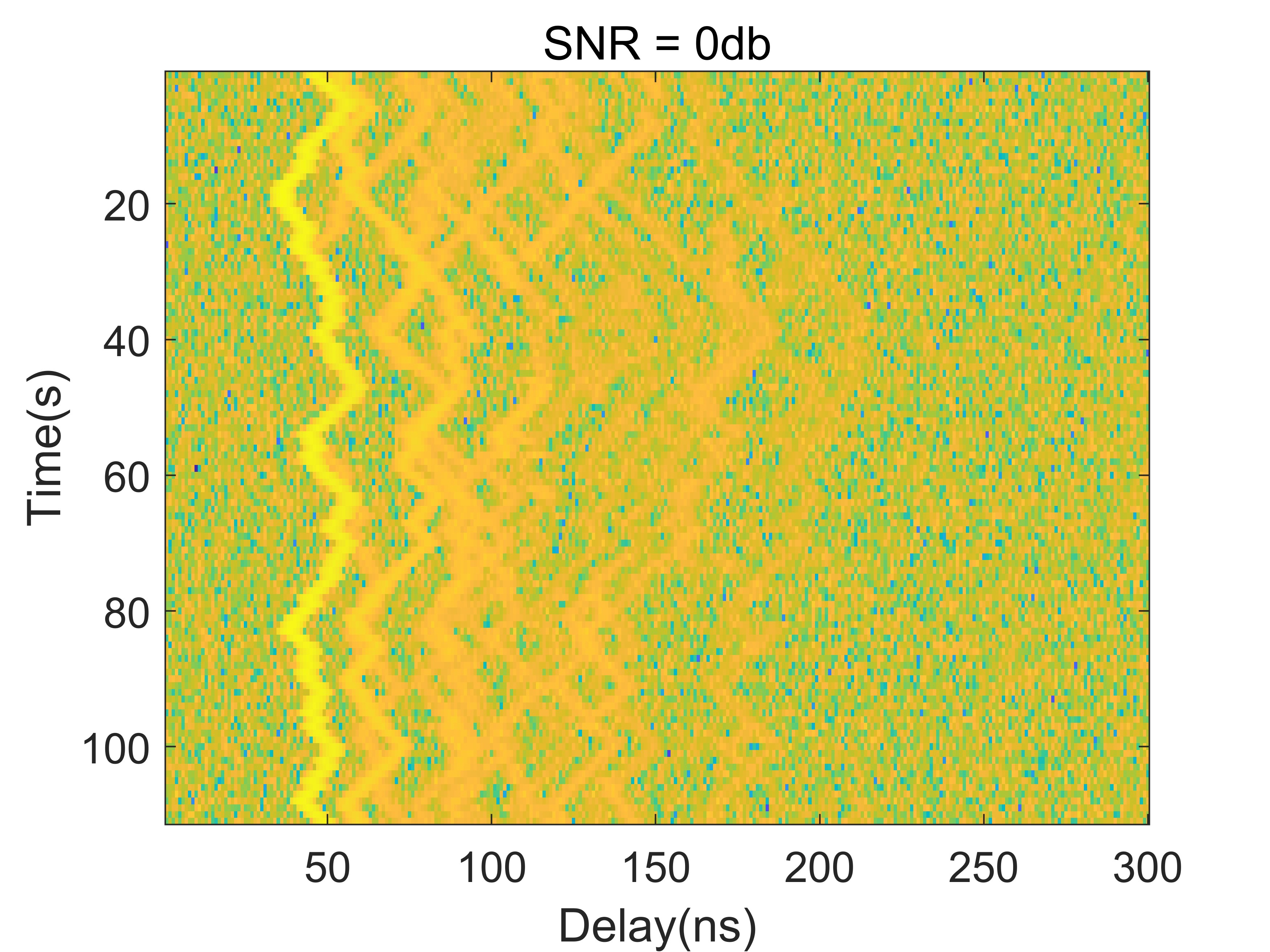}
        \end{minipage}
        \hspace{-0.5cm}
        \begin{minipage}{0.26\textwidth}
            \centering
            \includegraphics[width=\textwidth]{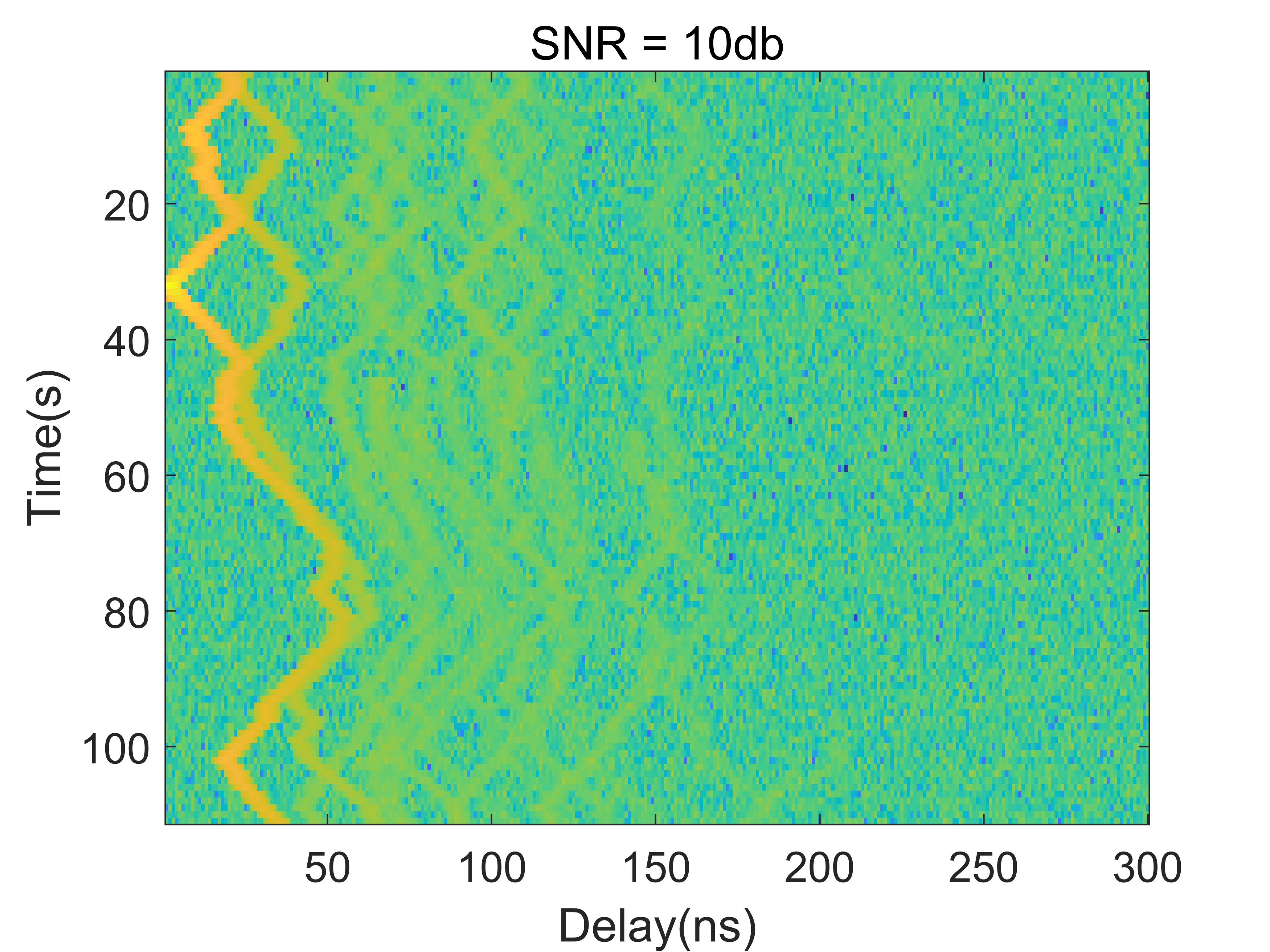}
        \end{minipage}
        \hspace{-0.5cm}
        \begin{minipage}{0.26\textwidth}
            \centering
            \includegraphics[width=\textwidth]{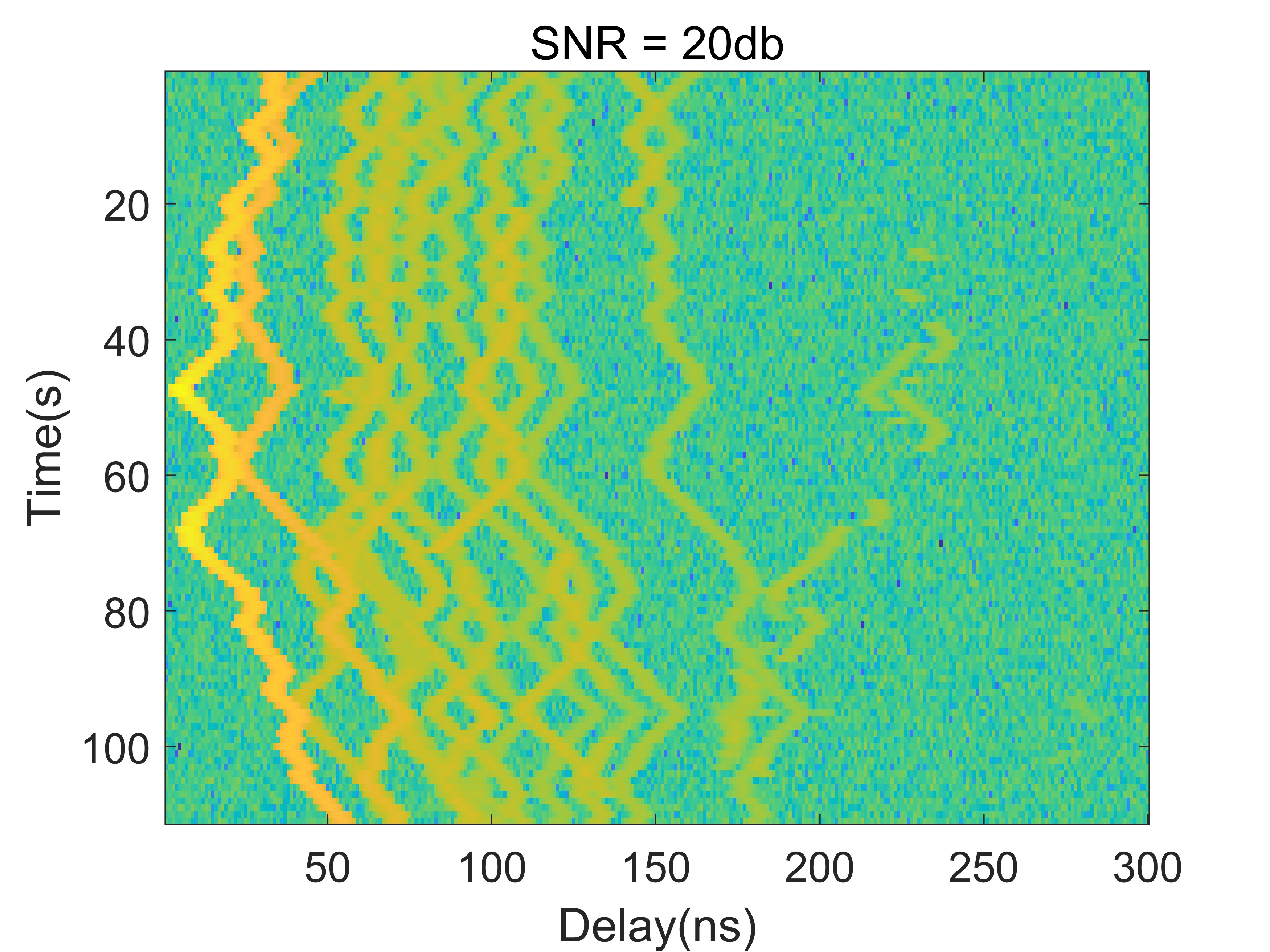}
        \end{minipage}
        \hspace{-0.5cm}
    \end{minipage}

    \vspace{0.2cm} 

    \begin{minipage}{\textwidth}
        \centering
        \begin{minipage}{0.26\textwidth}
            \centering
            \includegraphics[width=\textwidth]{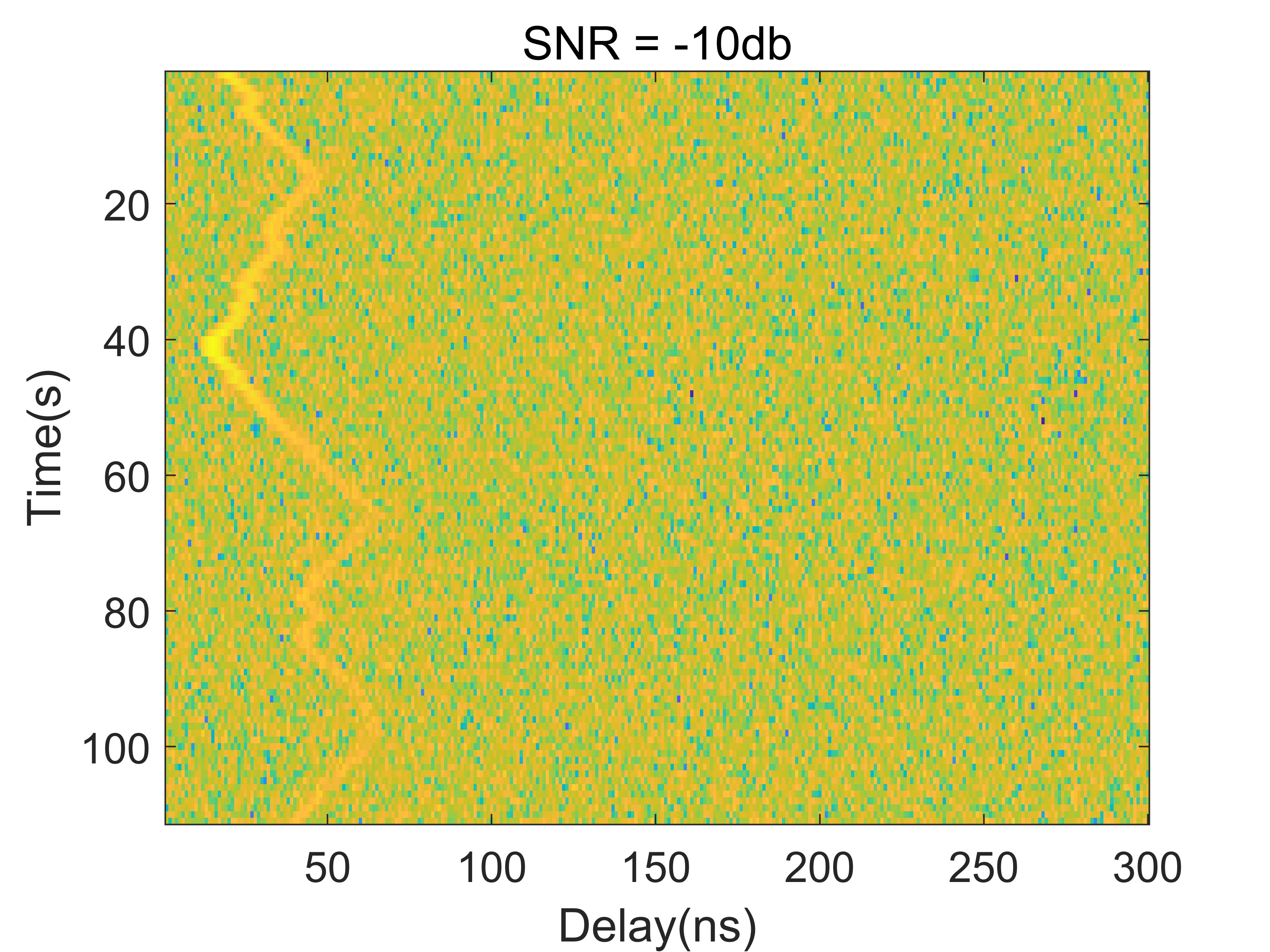}
        \end{minipage}
        \hspace{-0.5cm}
        \begin{minipage}{0.26\textwidth}
            \centering
            \includegraphics[width=\textwidth]{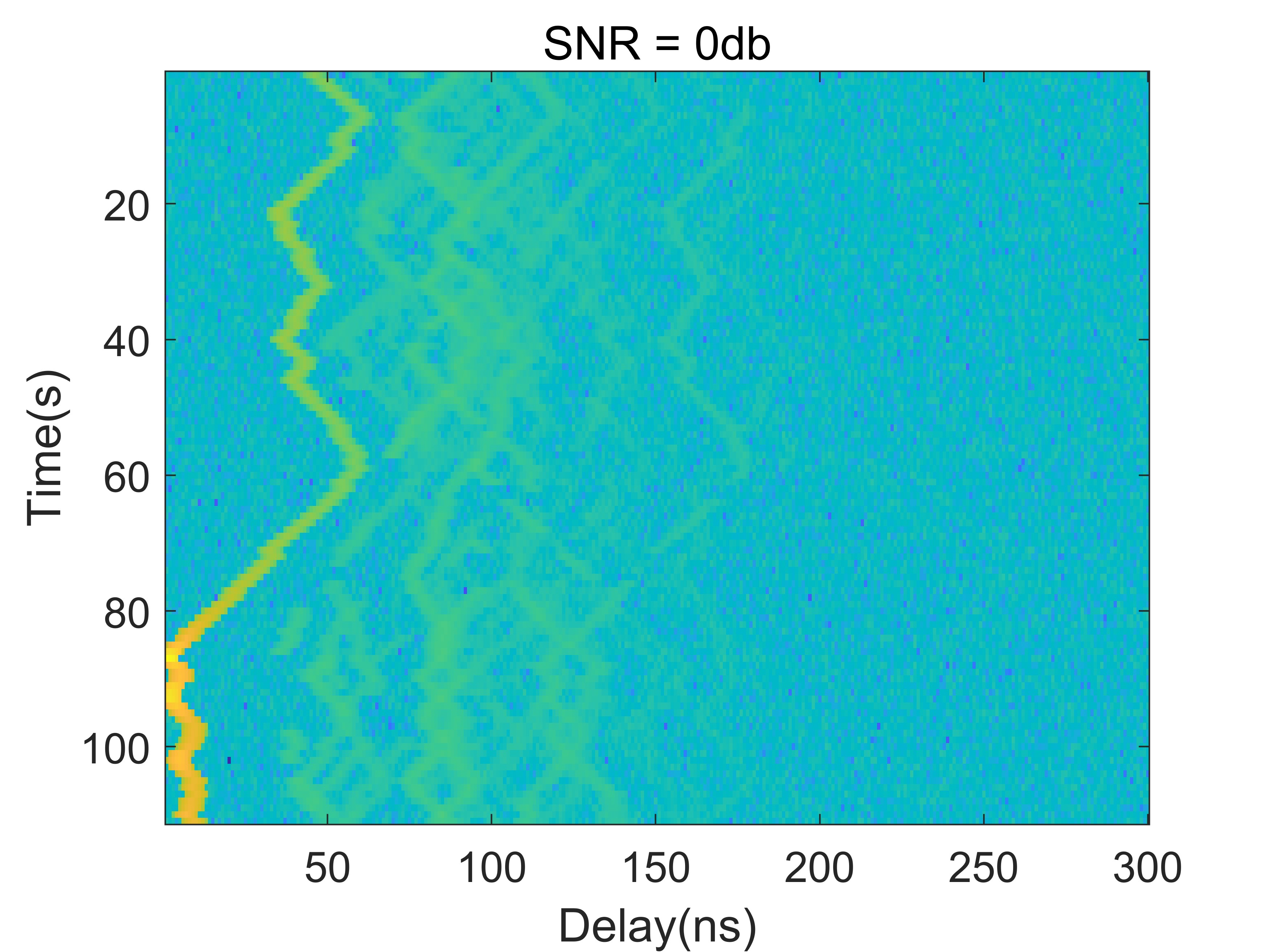}
        \end{minipage}
        \hspace{-0.5cm}
        \begin{minipage}{0.26\textwidth}
            \centering
            \includegraphics[width=\textwidth]{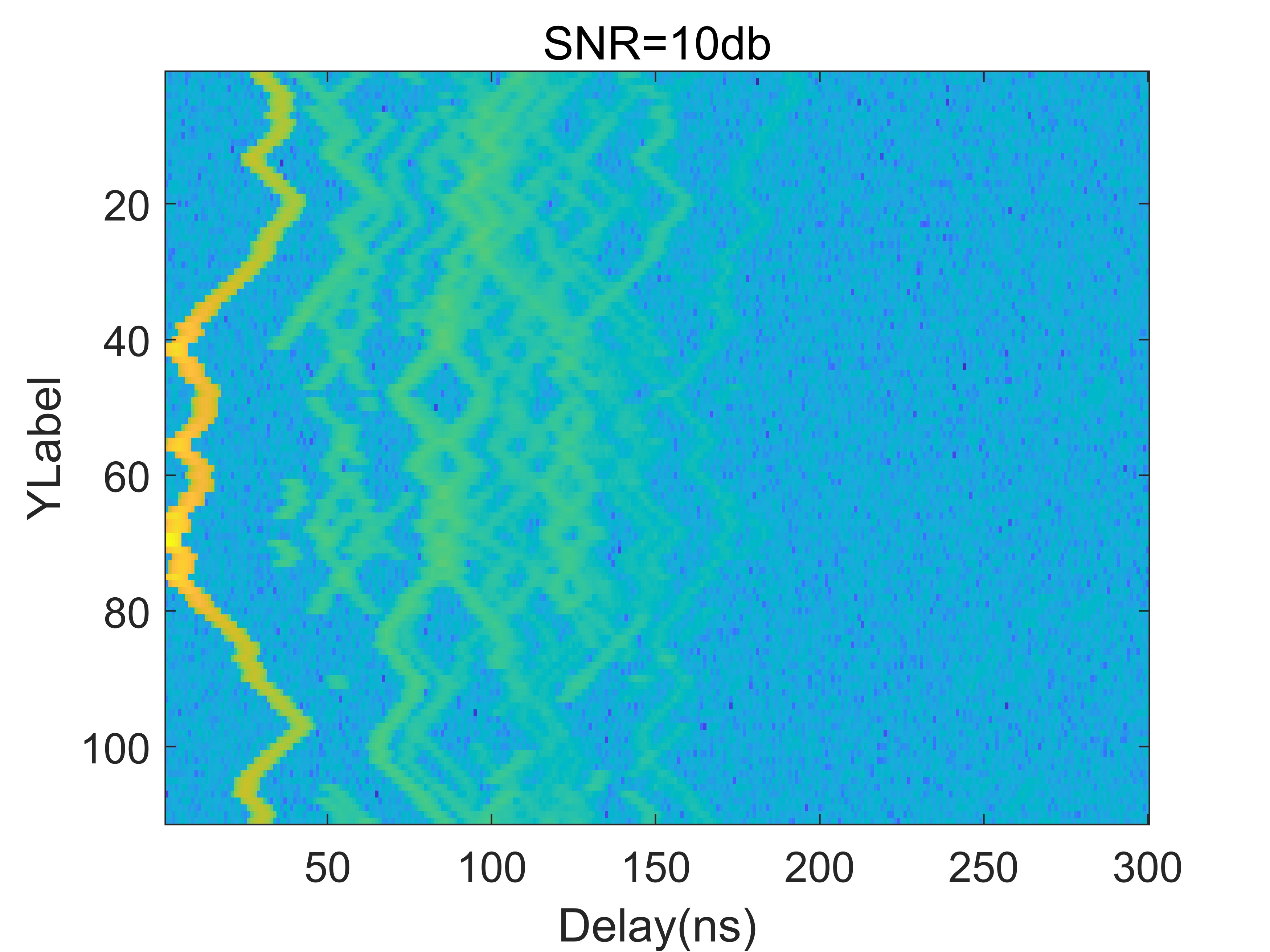}
        \end{minipage}
        \hspace{-0.5cm}
        \begin{minipage}{0.26\textwidth}
            \centering
            \includegraphics[width=\textwidth]{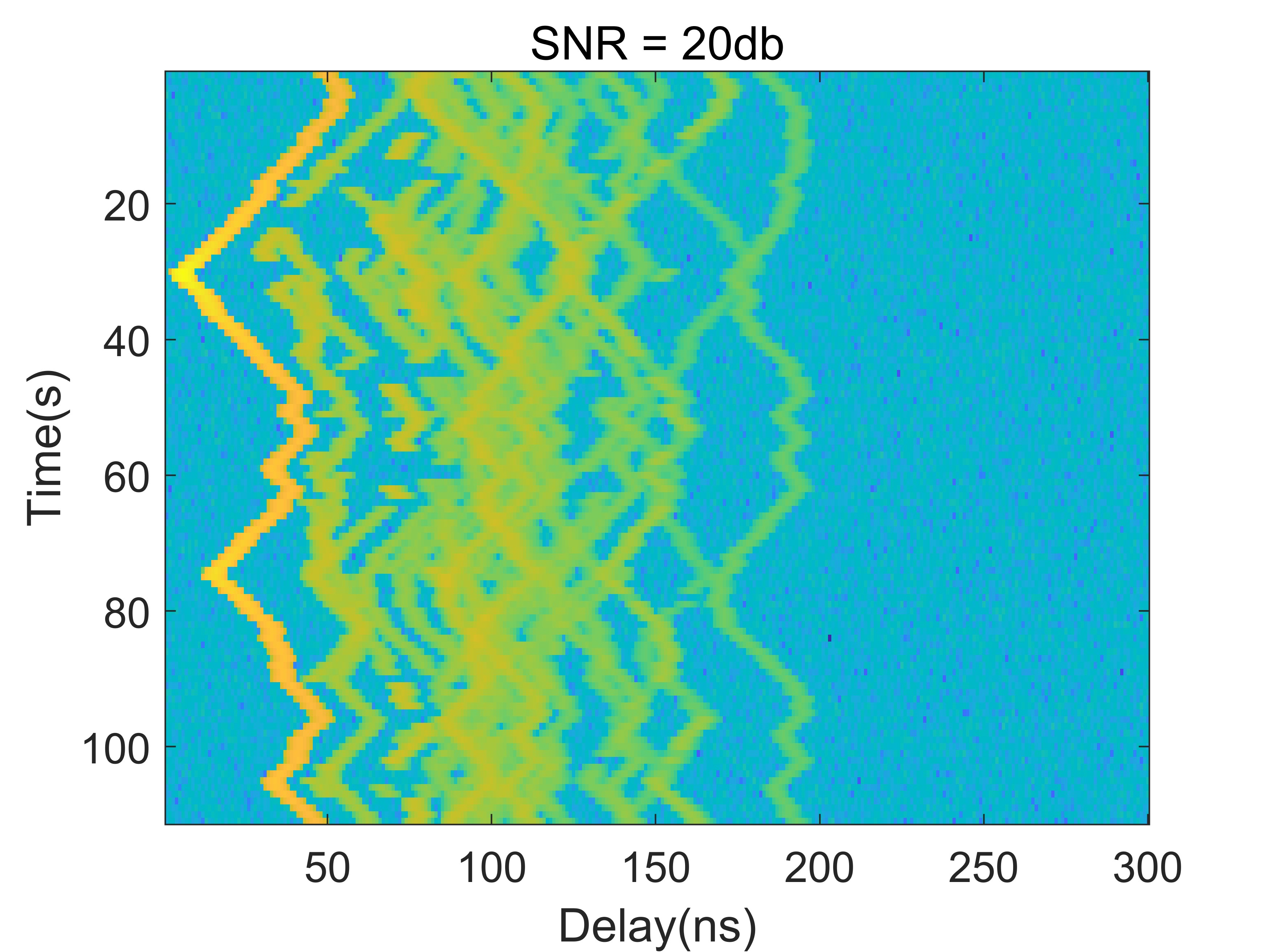}
        \end{minipage}
        \hspace{-0.5cm}
    \end{minipage}
    \caption{Two different methods for controlling SNR under bandwidth of 100Mhz: by adding additive Gaussian white noise (Top), and by controlling transmission power (Bottom). At higher SNR levels, the differences between the two methods are minimal, while at lower SNR levels, the transmission power control method demonstrates more stringent signal conditions. Especially when the SNR drops to -10 dB, only the \ac{fp} is visible. Compared to adding additive background noise, adjusting the SNR by controlling the transmission power makes the power levels of \ac{mpc}s at different orders closer under low SNR conditions, better aligning with the local multipath aliasing caused by diffuse reflection.}
    \label{fig:SNR}
\end{figure*}

We assume that the observation noise should satisfy the \ac{wss} assumption in the same indoor environment. Therefore, for each trajectory, we perform dynamic power allocation based on the 5G signaling communication mechanism \cite{lin2022overview} to control the \ac{snr} as closely as possible to the real situation, rather than adjusting the noise level. We use the \ac{fp} power at the start time $P_{FP}(t=0)$ of each trajectory as the signal power and calculate the \ac{snr} with the noise power being the background noise $P_{BN}$.
\begin{equation}
    SNR=10log_{10}\frac{P_{FP}(t=0)}{P_{BN}} - G_{ss}
\end{equation}
If additive noise is used, the power differences between multipath components with similar delays remain unchanged. However, adjusting the SNR by controlling the transmission power brings the power levels of different-order multipath components closer under low SNR conditions. In other words, this approach generates a series of aliasing multipath components with more similar power and delay in the multi-time CIR spectrum, better simulated the multipath aliasing caused by randomly distributed diffuse reflectors in a local region. We present the differences in multi-time \ac{cir}s under different \ac{snr} control methods for \ac{snr} levels of -10 dB, 0 dB, 10 dB, and 20 dB. Compared to adjusting the background noise level, controlling the \ac{snr} by adjusting the transmission power based on the first-path power as a reference presents more stringent experimental conditions. As shown in the figure \ref{fig:SNR}, even considering a spreading gain of 20 dB, the transmission power control method exhibits more indistinguishable multipath at the same \ac{snr}.

\subsection{Baseline Methods for Evaluation}
The methods used for comparison mainly include deep learning-based approaches, such as Multi-Layer Perceptron (MLP)\cite{tolstikhin2021mlp}, Convolutional Neural Network (CNN)\cite{simonyan2014very}, Long Short-Term Memory (LSTM)\cite{graves2012long}, Deepfi\cite{wang2016csi}, and Transformer\cite{vaswani2017attention}. Unlike the proposed method, which splits multi-time CIR samples into single-time CIR samples as input, the baseline methods, including MLP, CNN, LSTM, and Deepfi, avoid introducing erroneous trajectory temporal information by doing so. This is because our trajectory is randomly generated, and the positions at consecutive times do not strictly follow a temporal relationship. In fact, single-time sample inputs have led to better performance for these baseline methods. For the Transformer, as one of the components of our method, we used the same input as in the proposed approach.

In terms of parameter settings, to ensure a fair comparison, we adopted the same number of layers for the baseline methods as in the proposed method, particularly for MLP and Transformer, where we used the exact same number of layers and neurons per layer as in the proposed method.

Before being input into each method, these features are normalized to [0, 1]. During training, a batch size of 128 is consistently used, with a maximum of 300 epochs and a decaying learning rate:
\begin{equation}
    lr=0.0001 \times 0.9\exp^{epoch/2}+0.000005
\end{equation}
For all baseline methods, the loss function is defined in the equation (\ref{eq:mse}). All methods are optimized using the Adam optimizer\cite{kingma2014adam}.

\subsection{Performance of Different Bandwidth}
\begin{figure}
    \centering
    \includegraphics[width=1\linewidth]{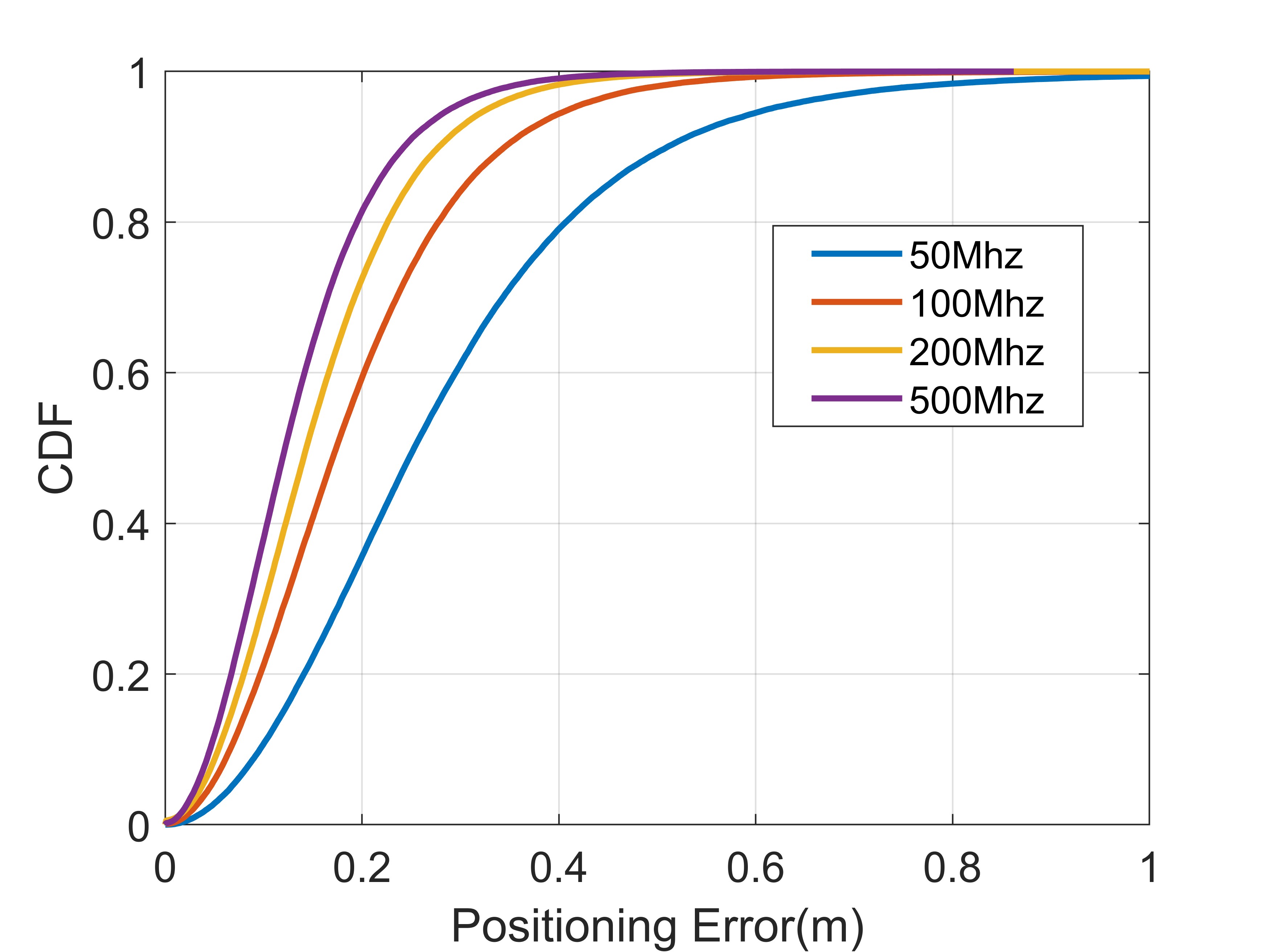}
    \caption{Error Cumulative Distribution Function for Different Bandwidth}
    \label{fig:bandwidth}
\end{figure}

Diffuse reflection is a major source of indistinguishable multipath, leading to the arrival of multiple random reflection paths with similar delays and powers at the receiver. This results in indistinguishable multipath in the received multi-time \ac{cir} observations. In this section, we represent different multipath resolutions through varying bandwidths to simulate the impact of diffuse reflection.
As shown in the Figure \ref{fig:bandwidth}, we evaluated the performance under different bandwidths in a noise-free condition. It can be observed that as the bandwidth increases, the multipath resolution improves, leading to performance enhancement. However, the amount of improvement diminishes as the bandwidth increases.
\subsection{Performance of Different Multipath Aliasing}

\begin{figure}[h]
    \centering
    \includegraphics[width=1\linewidth]{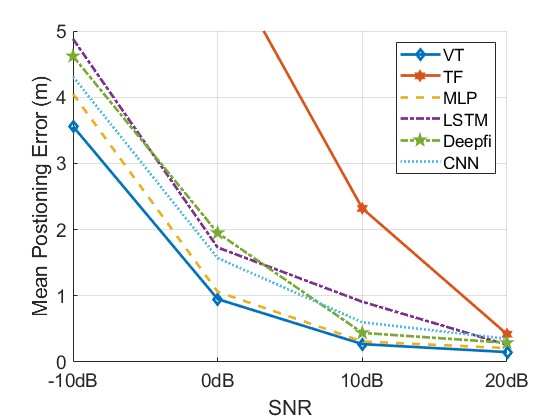}
    \caption{Mean Position Error under Different SNR Conditions}
    \label{fig:logerror}
\end{figure}

\begin{figure*}[!t]
\centering
\subfloat[Error Cumulative Distribution Function under SNR=-10dB]{{\includegraphics[width=0.5\textwidth]{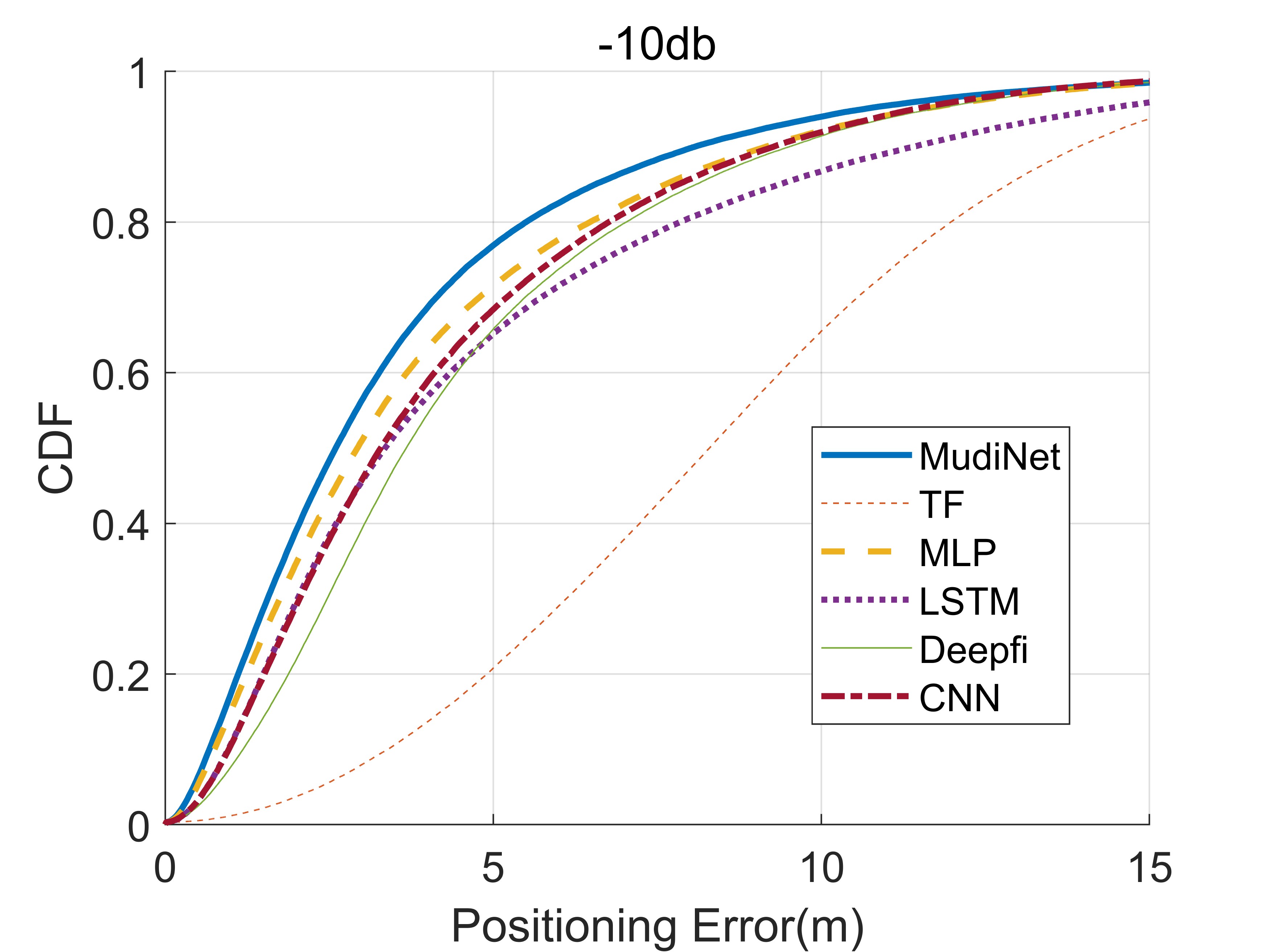}%
\label{fig:baseline_-10}}}
\hfil
\subfloat[Error Cumulative Distribution Function under SNR=0dB]{{\includegraphics[width=0.5\textwidth]{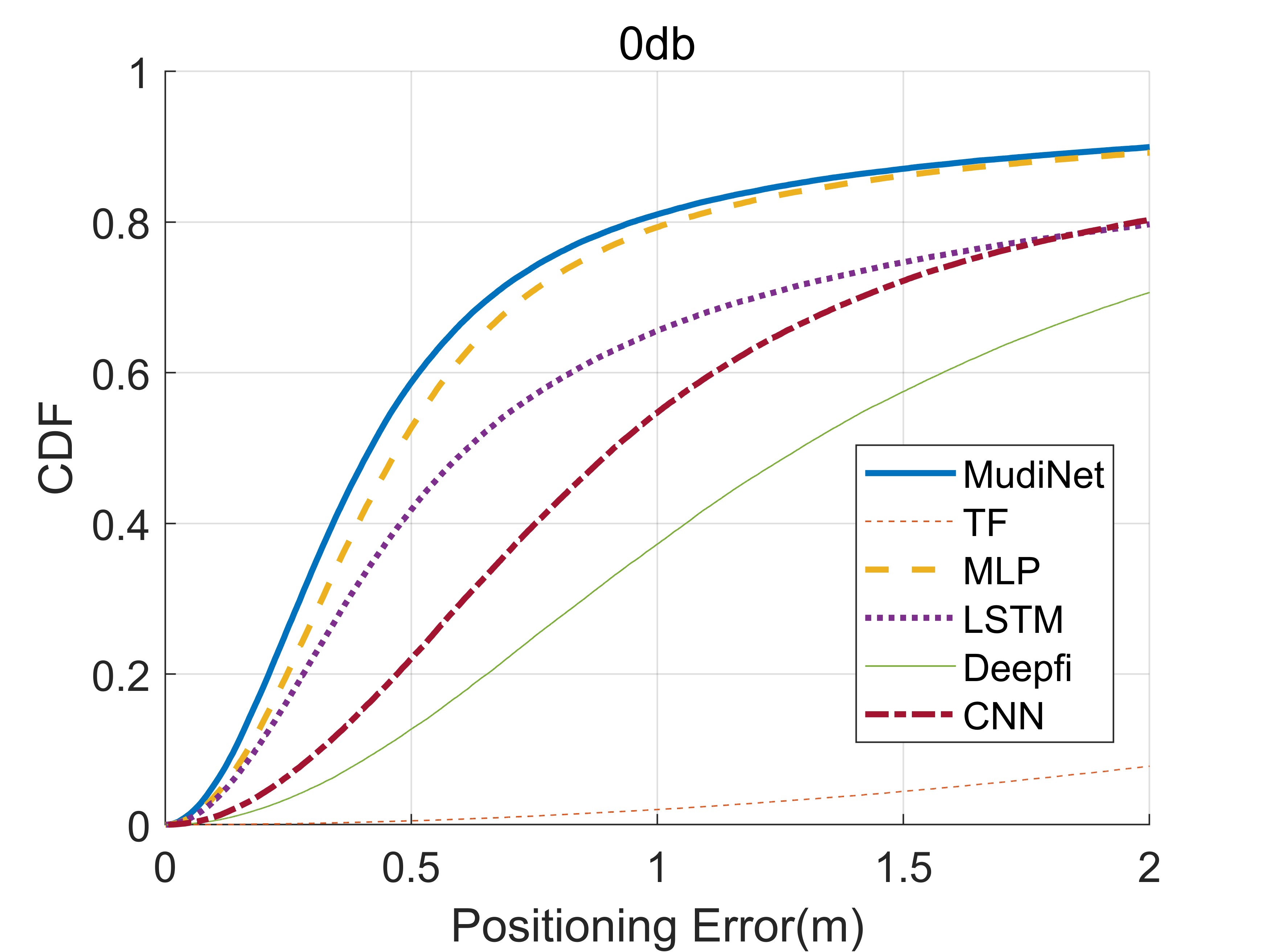}%
\label{fig:baseline_0}}}
\hfil
\\
\subfloat[Error Cumulative Distribution Function under SNR=10dB]{{\includegraphics[width=0.5\textwidth]{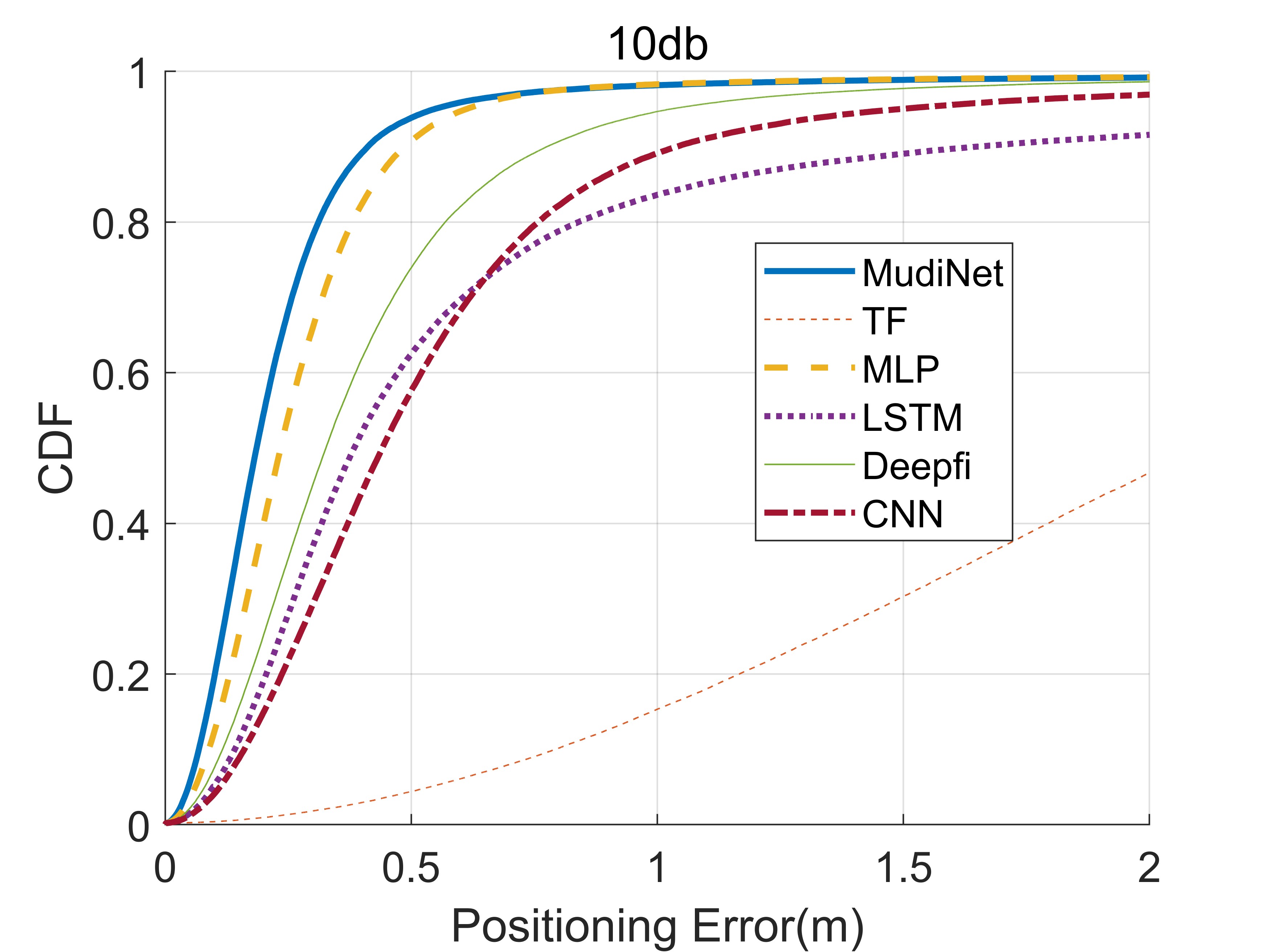}%
\label{fig:baseline_10}}}
\hfil
\subfloat[Error Cumulative Distribution Function under SNR=20dB]{{\includegraphics[width=0.5\textwidth]{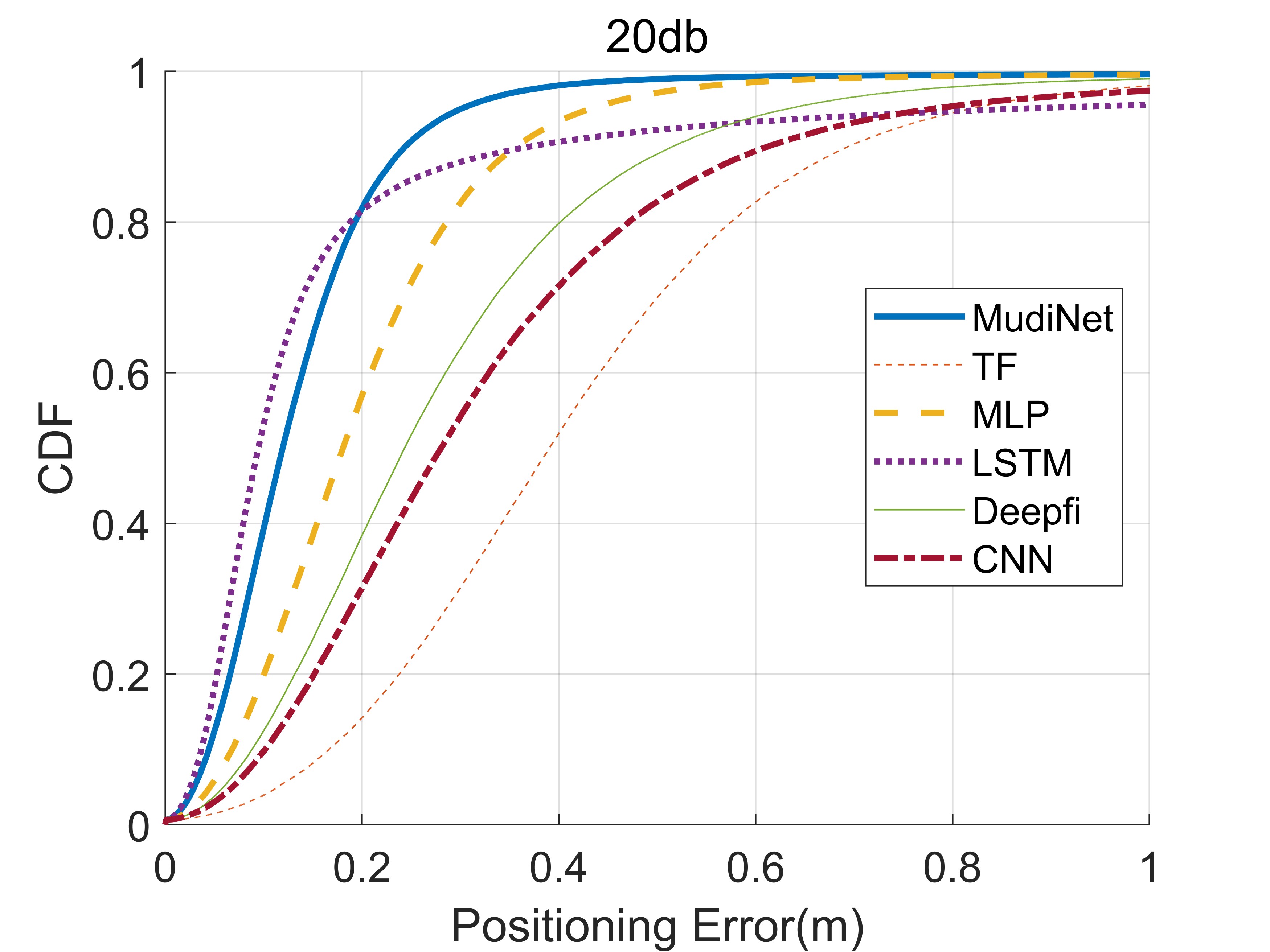}%
\label{fig:baseline_20}}}
\hfil
\caption{Improvements brought by the proposed method under different levels of multipath aliasing are compared,  and the degree of multipath aliasing is quantified using SNR. Various methods performed differently under different SNR levels, but our method consistently achieved the best performance.}
\label{fig:baseline}
\end{figure*}

In this section, we compare the performance differences between our method and other methods under various multipath interference conditions, where the degree of multipath interference is quantified by SNR. Table \ref{tab:acc} presents the mean error (ME) and \ac{rmse} of different methods, calculated based on all trajectories in the test set. Overall, the proposed method achieves the highest performance across different SNR levels. Notably, at SNR = 20 dB, it attains an average accuracy of approximately 0.15m and an RMSE of 0.32m. Figure \ref{fig:logerror} provides a more intuitive representation. For the MLP method, which maintains the same parameters as our approach, this serves as a fundamental ablation study. As SNR decreases, the proposed complete method MudiNet consistently exhibits a growing performance improvement over the ablated MLP method. In contrast, other methods demonstrate varying performance trends across different SNR conditions. This indicates that when multipath quality is high, our method can achieve high accuracy while maintaining good error stability, which can also be observed in Figure \ref{fig:baseline_20}. Under the premise of a smaller mean error, the proposed method exhibits smaller errors at cumulative probabilities exceeding 80\%.

Among the baseline methods, MLP achieves performance closest to our method. This is primarily because MLP adopts exactly the same parameter settings as our method but lacks the environmental variable estimation and separation components. Additionally, the difference in training methods plays a role. Our method incorporates multi-time CIR inputs for global environmental variable estimation. However, for a pure localization task, splitting multi-time CIR into single-time samples aligns better with standard localization frameworks. This is especially relevant given that our simulation employs randomly generated trajectories, where single-time samples prevent the model from incorrectly learning non-existent global position relationships. Furthermore, we observe that the Transformer with multi-time CIR inputs yields the worst results, supporting our argument that global information exists only in the estimation of environmental variables from multi-time CIR samples and is unrelated to the UE's position itself.

Among these baseline methods trained on single-time CIR samples, despite using the same number of layers, CNN and LSTM perform worse than MLP. This is because the sequential and power features along the delay axis contain redundant information. Specifically, in the delay axis, lower-power features inevitably appear to the right of higher-power features. Models like LSTM, which focus on sequential features, are more susceptible to such redundancy, leading to performance degradation. Meanwhile, CNN is essentially a sparsely connected MLP. Compared to MLP, CNN generally excels in image processing tasks due to its ability to preserve local spatial structures through 2D convolution kernels. However, in 1D data processing, this unique advantage is lost.

More importantly, we observe that as SNR decreases and the proportion of indistinguishable multipath increases, our proposed method demonstrates greater performance improvements. At SNR = 20 dB, our method achieves only a 0.06m improvement over the second-best method and a 0.11m improvement over the third-best method. However, at SNR = -10 dB, the improvement increases to 0.49m and 0.75m, respectively. For the second-best and third-best methods, at 20 dB, the second-best method improves by 0.05m compared to the third-best method, whereas at -10 dB, this improvement is only 0.26m. This aligns with our hypothesis that the proposed method is more effective in mitigating the impact of indistinguishable multipath.

Further examination of the error distribution, as illustrated in Figure \ref{fig:baseline}, reveals that the baseline methods based on single-time CIR samples display varying error magnitudes at different cumulative probabilities. In contrast, our method exhibits a clear advantage under low SNR conditions, achieving smaller errors across all cumulative probability levels. As shown in Figures \ref{fig:baseline}, this advantage diminishes as the SNR increases, but in general, it still maintains its advantages.

\begin{table*}[h]
    \centering
    \renewcommand\arraystretch{1.2}
    \caption{The Mean Error and Root Mean Squared Error of different methods under different SNRs. The best performance is marked in \red{red} and bold, the second-best in \green{green} and bold, and the third-best in \blue{blue} and bold.}
    \begin{tabular}{cccccc}
    \toprule
        \textbf{Methods} & \textbf{Metrics} &\textbf{SNR=-10db} &\textbf{SNR=0db} & \textbf{SNR=10db} & \textbf{SNR=20db}\\
        \midrule
        \multirow{2}{*}{MudiNet(proposed)} & ME(m) &\red{\textbf{3.56}} & \red{\textbf{0.95}} & \red{\textbf{0.27}} & \red{\textbf{0.15}} \\
        & RMSE(m) & 4.95 & 2.10 & 0.66 & 0.32\\
        \hline
        \multirow{2}{*}{TF} & ME(m) &8.51 & 6.25 & 2.32 & 0.42 \\
        & RMSE(m) & 9.41 & 7.10 & 2.72 & 0.55\\
        \hline
        \multirow{2}{*}{MLP} & ME(m) &\green{\textbf{4.05}} & \green{\textbf{1.06}} & \green{\textbf{0.31}} & \green{\textbf{0.21}} \\
        & RMSE(m) & 5.40 & 2.26 & 0.67 & 0.35\\
        \hline
        \multirow{2}{*}{LSTM} & ME(m) &4.88 & 1.73 & 0.91 & \blue{\textbf{0.26}} \\
        & RMSE(m) & 6.53 & 3.41 & 2.00 & 0.87\\
        \hline
        \multirow{2}{*}{Deepfi} & ME(m) &4.62 & 1.95 & \blue{\textbf{0.44}} & 0.29 \\
        & RMSE(m) & 5.76 & 2.94 & 0.78 & 0.44\\
        \hline
        \multirow{2}{*}{CNN} & ME(m) &\blue{\textbf{4.31}} & \blue{\textbf{1.57}} & 0.60 & 0.35 \\
        & RMSE(m) & 5.52 & 2.55 & 0.96 & 0.55\\
      
    \bottomrule
    \end{tabular}
    \label{tab:acc}
\end{table*}

\section{Conclusion}
In this paper, we model diffuse multipath as an indistinguishable multipath component caused by the random distribution of diffuse scatterers in physical space and demonstrate that indistinguishable multipath is statistically independent and separable from ideal specular multipath, which positively contributes to localization. Based on this understanding, we design a deep learning method based on variational inference to enhance localization accuracy by separating and suppressing indistinguishable multipath, particularly improving localization robustness under severe indistinguishable multipath conditions.

We conducted simulation experiments in a \ac{siso} single-base-station localization scenario, where dynamic transmission power adjustment was used to simulate multipath aliasing under different SNR conditions. Compared to the baseline methods, as the SNR decreased, our method's performance improvement over the second-best method increased from 0.06m to 0.49m, while the second-best method's improvement over the third-best method increased from 0.05m to 0.26m. This demonstrates the effectiveness of our approach in enhancing localization accuracy and robustness under severe multipath aliasing conditions.

However, this work is just the beginning, providing insight into multipath separation. Our future research will focus on multipath channel sensing and partial CIR reconstruction to further improve localization performance and support downstream applications. Additionally, our method has been evaluated based on simulated measurements, necessitating further real-world empirical studies.

\section*{acknowledgment}
We sincerely thank Mrs. Zhou for the schematic design. This work has been submitted to the IEEE for possible publication. Copyright may be transferred without notice, after which this version may no longer be accessible.


\bibliographystyle{IEEEtran}
\bibliography{IEEEabrv, ref}

\begin{thebibliography}{10}
\providecommand{\url}[1]{#1}
\csname url@samestyle\endcsname
\providecommand{\newblock}{\relax}
\providecommand{\bibinfo}[2]{#2}
\providecommand{\BIBentrySTDinterwordspacing}{\spaceskip=0pt\relax}
\providecommand{\BIBentryALTinterwordstretchfactor}{4}
\providecommand{\BIBentryALTinterwordspacing}{\spaceskip=\fontdimen2\font plus
\BIBentryALTinterwordstretchfactor\fontdimen3\font minus \fontdimen4\font\relax}
\providecommand{\BIBforeignlanguage}[2]{{%
\expandafter\ifx\csname l@#1\endcsname\relax
\typeout{** WARNING: IEEEtran.bst: No hyphenation pattern has been}%
\typeout{** loaded for the language `#1'. Using the pattern for}%
\typeout{** the default language instead.}%
\else
\language=\csname l@#1\endcsname
\fi
#2}}
\providecommand{\BIBdecl}{\relax}
\BIBdecl

\bibitem{shao2020accurate}
W.~Shao, H.~Luo, F.~Zhao, H.~Tian, S.~Yan, and A.~Crivello, ``Accurate indoor positioning using temporal--spatial constraints based on {Wi-Fi} fine time measurements,'' \emph{IEEE Internet of Things Journal}, vol.~7, no.~11, pp. 11\,006--11\,019, 2020.

\bibitem{yu2021novel}
Y.~Yu, R.~Chen, L.~Chen, X.~Zheng, D.~Wu, W.~Li, and Y.~Wu, ``A novel {3-D} indoor localization algorithm based on {BLE} and multiple sensors,'' \emph{IEEE Internet of Things Journal}, vol.~8, no.~11, pp. 9359--9372, 2021.

\bibitem{wang2023recent}
Q.~Wang, M.~Fu, J.~Wang, H.~Luo, L.~Sun, Z.~Ma, W.~Li, C.~Zhang, R.~Huang, X.~Li \emph{et~al.}, ``Recent advances in floor positioning based on smartphone,'' \emph{Measurement}, vol. 214, p. 112813, 2023.

\bibitem{zhao2021new}
S.~Zhao, X.-P. Zhang, X.~Cui, and M.~Lu, ``A new {TOA} localization and synchronization system with virtually synchronized periodic asymmetric ranging network,'' \emph{IEEE Internet of Things Journal}, vol.~8, no.~11, pp. 9030--9044, 2021.

\bibitem{liu2022integrated}
F.~Liu, Y.~Cui, C.~Masouros, J.~Xu, T.~X. Han, Y.~C. Eldar, and S.~Buzzi, ``Integrated sensing and communications: Toward dual-functional wireless networks for {6G} and beyond,'' \emph{IEEE journal on selected areas in communications}, vol.~40, no.~6, pp. 1728--1767, 2022.

\bibitem{del2017survey}
J.~A. del Peral-Rosado, R.~Raulefs, J.~A. L{\'o}pez-Salcedo, and G.~Seco-Granados, ``Survey of cellular mobile radio localization methods: From {1G} to {5G},'' \emph{IEEE Communications Surveys \& Tutorials}, vol.~20, no.~2, pp. 1124--1148, 2017.

\bibitem{multipathsurvey2024}
L.~Xie, Y.~Zhang, C.~Zhao, C.~Zhang, Z.~Mu, and X.~Yang, ``Positioning under multipath environments in wireless network: Survey, design and opportunities,'' \emph{IEEE Network}, 2024.

\bibitem{gentner2016multipath}
C.~Gentner, T.~Jost, W.~Wang, S.~Zhang, A.~Dammann, and U.-C. Fiebig, ``Multipath assisted positioning with simultaneous localization and mapping,'' \emph{IEEE Transactions on Wireless Communications}, vol.~15, no.~9, pp. 6104--6117, 2016.

\bibitem{ge2021single}
F.~Ge and Y.~Shen, ``Single-anchor ultra-wideband localization system using wrapped {PDoA},'' \emph{IEEE Transactions on Mobile Computing}, vol.~21, no.~12, pp. 4609--4623, 2021.

\bibitem{nazari2023mmwave}
M.~A. Nazari, G.~Seco-Granados, P.~Johannisson, and H.~Wymeersch, ``Mmwave {6D} radio localization with a snapshot observation from a single {BS},'' \emph{{IEEE} Trans. Veh. Technol.}, vol.~72, no.~7, pp. 8914--8928, Jul. 2023.

\bibitem{liu2023multipath}
Z.~Liu, L.~Chen, X.~Zhou, N.~Shen, and R.~Chen, ``Multipath tracking with {LTE} signals for accurate {TOA} estimation in the application of indoor positioning,'' \emph{Geo-spatial Information Science}, vol.~26, no.~1, pp. 31--43, 2023.

\bibitem{liu2023machine}
Z.~Liu, L.~Chen, X.~Zhou, Z.~Jiao, G.~Guo, and R.~Chen, ``Machine learning for time-of-arrival estimation with {5G} signals in indoor positioning,'' \emph{IEEE Internet of Things Journal}, vol.~10, no.~11, pp. 9782--9795, 2023.

\bibitem{zhou2022aoa}
T.~Zhou, K.~Xu, Z.~Shen, W.~Xie, D.~Zhang, and J.~Xu, ``{AoA}-based positioning for aerial intelligent reflecting surface-aided wireless communications: An angle-domain approach,'' \emph{IEEE Wireless Communications Letters}, vol.~11, no.~4, pp. 761--765, 2022.

\bibitem{soltanaghaei2018multipath}
E.~Soltanaghaei, A.~Kalyanaraman, and K.~Whitehouse, ``Multipath triangulation: Decimeter-level wifi localization and orientation with a single unaided receiver,'' in \emph{Proceedings of the 16th annual international conference on mobile systems, applications, and services}, 2018, pp. 376--388.

\bibitem{li2019massive}
X.~Li, E.~Leitinger, M.~Oskarsson, K.~{\AA}str{\"o}m, and F.~Tufvesson, ``Massive {MIMO}-based localization and mapping exploiting phase information of multipath components,'' \emph{IEEE transactions on wireless communications}, vol.~18, no.~9, pp. 4254--4267, 2019.

\bibitem{du2024diffuse}
Y.~Du, H.~Zhao, Y.~Liu, X.~Yu, and Y.~Shen, ``Exploiting multipath information for integrated localization and sensing via {PHD} filtering,'' \emph{IEEE Transactions on Vehicular Technology}, 2024.

\bibitem{kim2022uwb}
D.-H. Kim, A.~Farhad, and J.-Y. Pyun, ``{UWB} positioning system based on {LSTM} classification with mitigated {NLOS} effects,'' \emph{IEEE Internet of Things Journal}, vol.~10, no.~2, pp. 1822--1835, 2022.

\bibitem{wang2024multipath}
T.~Wang, Y.~Li, J.~Liu, K.~Hu, and Y.~Shen, ``Multipath-assisted single-anchor localization via deep variational learning,'' \emph{IEEE Transactions on Wireless Communications}, 2024.

\bibitem{li2023variational}
Y.~Li, S.~Mazuelas, and Y.~Shen, ``A variational learning approach for concurrent distance estimation and environmental identification,'' \emph{IEEE Transactions on Wireless Communications}, vol.~22, no.~9, pp. 6252--6266, 2023.

\bibitem{xu2023multi}
X.~Xu, A.~Peng, X.~Hong, Y.~Zhang, and X.-P. Zhang, ``Multi-state constraint multipath-assisted positioning and mismatch alleviation,'' \emph{IEEE Internet of Things Journal}, 2023.

\bibitem{tseng2017ray}
P.-H. Tseng, Y.-C. Chan, Y.-J. Lin, D.-B. Lin, N.~Wu, and T.-M. Wang, ``Ray-tracing-assisted fingerprinting based on channel impulse response measurement for indoor positioning,'' \emph{IEEE Transactions on Instrumentation and Measurement}, vol.~66, no.~5, pp. 1032--1045, 2017.

\bibitem{gao2022toward}
K.~Gao, H.~Wang, H.~Lv, and W.~Liu, ``Toward {5G} {NR} high-precision indoor positioning via channel frequency response: A new paradigm and dataset generation method,'' \emph{IEEE Journal on Selected Areas in Communications}, vol.~40, no.~7, pp. 2233--2247, 2022.

\bibitem{wang2016csi}
X.~Wang, L.~Gao, S.~Mao, and S.~Pandey, ``{CSI}-based fingerprinting for indoor localization: A deep learning approach,'' \emph{IEEE transactions on vehicular technology}, vol.~66, no.~1, pp. 763--776, 2016.

\bibitem{ruan2022ipos}
Y.~Ruan, L.~Chen, X.~Zhou, Z.~Liu, X.~Liu, G.~Guo, and R.~Chen, ``{iPos-5G}: Indoor positioning via commercial {5G NR CSI},'' \emph{IEEE Internet of Things Journal}, vol.~10, no.~10, pp. 8718--8733, 2022.

\bibitem{ruan2022hi}
Y.~Ruan, L.~Chen, X.~Zhou, G.~Guo, and R.~Chen, ``Hi-loc: Hybrid indoor localization via enhanced {5G NR CSI},'' \emph{IEEE Transactions on Instrumentation and Measurement}, vol.~71, pp. 1--15, 2022.

\bibitem{kram2019uwb}
S.~Kram, M.~Stahlke, T.~Feigl, J.~Seitz, and J.~Thielecke, ``{UWB} channel impulse responses for positioning in complex environments: A detailed feature analysis,'' \emph{Sensors}, vol.~19, no.~24, p. 5547, 2019.

\bibitem{chen2020uwb}
Y.-Y. Chen, S.-P. Huang, T.-W. Wu, W.-T. Tsai, C.-Y. Liou, and S.-G. Mao, ``{UWB} system for indoor positioning and tracking with arbitrary target orientation, optimal anchor location, and adaptive {NLOS} mitigation,'' \emph{IEEE Transactions on Vehicular Technology}, vol.~69, no.~9, pp. 9304--9314, 2020.

\bibitem{witrisal2016high}
K.~Witrisal, P.~Meissner, E.~Leitinger, Y.~Shen, C.~Gustafson, F.~Tufvesson, K.~Haneda, D.~Dardari, A.~F. Molisch, A.~Conti \emph{et~al.}, ``High-accuracy localization for assisted living: {5G} systems will turn multipath channels from foe to friend,'' \emph{IEEE Signal Processing Magazine}, vol.~33, no.~2, pp. 59--70, 2016.

\bibitem{leitinger2015evaluation}
E.~Leitinger, P.~Meissner, C.~R{\"u}disser, G.~Dumphart, and K.~Witrisal, ``Evaluation of position-related information in multipath components for indoor positioning,'' \emph{IEEE Journal on Selected Areas in communications}, vol.~33, no.~11, pp. 2313--2328, 2015.

\bibitem{wielandner2023multipath}
L.~Wielandner, A.~Venus, T.~Wilding, and E.~Leitinger, ``Multipath-based {SLAM} for non-ideal reflective surfaces exploiting multiple-measurement data association,'' \emph{arXiv preprint arXiv:2304.05680}, 2023.

\bibitem{wilding2018accuracy}
T.~Wilding, S.~Grebien, U.~M{\"u}hlmann, and K.~Witrisal, ``Accuracy bounds for array-based positioning in dense multipath channels,'' \emph{Sensors}, vol.~18, no.~12, p. 4249, 2018.

\bibitem{guo2024angle}
L.~Guo, T.~Lv, and J.~Zeng, ``Angle-based positioning estimation leveraging diffuse scattering paths in millimeter-wave {MIMO} systems,'' \emph{IEEE Sensors Journal}, 2024.

\bibitem{wen20205g}
F.~Wen, J.~Kulmer, K.~Witrisal, and H.~Wymeersch, ``{5G} positioning and mapping with diffuse multipath,'' \emph{IEEE Transactions on Wireless Communications}, vol.~20, no.~2, pp. 1164--1174, 2020.

\bibitem{ge20205g}
Y.~Ge, F.~Wen, H.~Kim, M.~Zhu, F.~Jiang, S.~Kim, L.~Svensson, and H.~Wymeersch, ``{5G} {SLAM} using the clustering and assignment approach with diffuse multipath,'' \emph{Sensors}, vol.~20, no.~16, p. 4656, 2020.

\bibitem{jiang20193}
H.~Jiang, Z.~Zhang, L.~Wu, J.~Dang, and G.~Gui, ``A {3-D} non-stationary wideband geometry-based channel model for {MIMO} vehicle-to-vehicle communications in tunnel environments,'' \emph{IEEE Transactions on Vehicular Technology}, vol.~68, no.~7, pp. 6257--6271, 2019.

\bibitem{cheng2022channel}
X.~Cheng, Z.~Huang, and L.~Bai, ``Channel nonstationarity and consistency for beyond {5G} and {6G}: A survey,'' \emph{IEEE Communications Surveys \& Tutorials}, vol.~24, no.~3, pp. 1634--1669, 2022.

\bibitem{bai20213}
L.~Bai, Z.~Huang, H.~Du, and X.~Cheng, ``A {3-D} nonstationary wideband {V2V GBSM} with {UPAs} for massive {MIMO} wireless communication systems,'' \emph{IEEE Internet of Things Journal}, vol.~8, no.~24, pp. 17\,622--17\,638, 2021.

\bibitem{zhang2019learning}
Y.-J. Zhang, S.~Pan, L.~He, and Z.-H. Ling, ``Learning latent representations for style control and transfer in end-to-end speech synthesis,'' in \emph{ICASSP 2019-2019 IEEE International Conference on Acoustics, Speech and Signal Processing (ICASSP)}.\hskip 1em plus 0.5em minus 0.4em\relax IEEE, 2019, pp. 6945--6949.

\bibitem{cifka2021self}
O.~C{\'\i}fka, A.~Ozerov, U.~{\c{S}}im{\c{s}}ekli, and G.~Richard, ``Self-supervised vq-vae for one-shot music style transfer,'' in \emph{ICASSP 2021-2021 IEEE International Conference on Acoustics, Speech and Signal Processing (ICASSP)}.\hskip 1em plus 0.5em minus 0.4em\relax IEEE, 2021, pp. 96--100.

\bibitem{liu2021multiple}
Z.-S. Liu, V.~Kalogeiton, and M.-P. Cani, ``Multiple style transfer via variational autoencoder,'' in \emph{2021 IEEE International Conference on Image Processing (ICIP)}.\hskip 1em plus 0.5em minus 0.4em\relax IEEE, 2021, pp. 2413--2417.

\bibitem{leitinger2019belief}
E.~Leitinger, F.~Meyer, F.~Hlawatsch, K.~Witrisal, F.~Tufvesson, and M.~Z. Win, ``A belief propagation algorithm for multipath-based {SLAM},'' \emph{IEEE transactions on wireless communications}, vol.~18, no.~12, pp. 5613--5629, 2019.

\bibitem{amiri2023indoor}
R.~Amiri, S.~Yerramalli, T.~Yoo, M.~Hirzallah, M.~Zorgui, R.~Prakash, and X.~Zhang, ``Indoor environment learning via {RF}-mapping,'' \emph{IEEE Journal on Selected Areas in Communications}, vol.~41, no.~6, pp. 1859--1872, 2023.

\bibitem{vaswani2017attention}
A.~Vaswani, ``Attention is all you need,'' \emph{Advances in Neural Information Processing Systems}, 2017.

\bibitem{kingma2013auto}
D.~P. Kingma, ``Auto-encoding variational bayes,'' \emph{arXiv preprint arXiv:1312.6114}, 2013.

\bibitem{lin2022overview}
X.~Lin, ``An overview of {5G} advanced evolution in {3GPP} release 18,'' \emph{IEEE Communications Standards Magazine}, vol.~6, no.~3, pp. 77--83, 2022.

\bibitem{tolstikhin2021mlp}
I.~O. Tolstikhin, N.~Houlsby, A.~Kolesnikov, L.~Beyer, X.~Zhai, T.~Unterthiner, J.~Yung, A.~Steiner, D.~Keysers, J.~Uszkoreit \emph{et~al.}, ``Mlp-mixer: An all-mlp architecture for vision,'' \emph{Advances in neural information processing systems}, vol.~34, pp. 24\,261--24\,272, 2021.

\bibitem{simonyan2014very}
K.~Simonyan, ``Very deep convolutional networks for large-scale image recognition,'' \emph{arXiv preprint arXiv:1409.1556}, 2014.

\bibitem{graves2012long}
A.~Graves and A.~Graves, ``Long short-term memory,'' \emph{Supervised sequence labelling with recurrent neural networks}, pp. 37--45, 2012.

\bibitem{kingma2014adam}
D.~P. Kingma, ``Adam: A method for stochastic optimization,'' \emph{arXiv preprint arXiv:1412.6980}, 2014.

\end{thebibliography}

\end{document}